\documentclass[preprint,12pt]{elsarticle}
\usepackage{color}
\definecolor{red}{rgb}{1,0,0}
\definecolor{green}{rgb}{0,1,0}
\definecolor{blue}{rgb}{0,0,1}

\usepackage{graphicx}
\usepackage{amssymb}
\usepackage{amsmath}
\usepackage{bm}

\usepackage{subcaption}
\usepackage{etoolbox}
\usepackage{tikz}

\newrobustcmd*{\mysquare}[1]{\tikz{\filldraw[draw=#1,fill=#1] (0,0)
rectangle (0.2cm,0.2cm);}}

\newrobustcmd*{\mycircle}[1]{\tikz{\filldraw[draw=#1,fill=#1] (0,0) circle [radius=0.1cm];}}

\newrobustcmd*{\mydowntriangle}[1]{\tikz{\filldraw[draw=#1,fill=#1] (0,0.2cm) --
(0.2cm,0.2cm) -- (0.1cm,0cm);}}

\newrobustcmd*{\myuptriangle}[1]{\tikz{\filldraw[draw=#1,fill=#1] (0,0) --
(0.2cm,0) -- (0.1cm,0.2cm);}}

\journal{International Journal of Multiphase Flow}

\begin{document}

\begin{frontmatter}



\title{Phase-field simulation of core-annular pipe flow}

\author[1,2]{Baofang Song}
\author[1]{Carlos Plana}
\author[3]{Jose M. Lopez}
\author[1]{Marc Avila\corref{cor1}}
\ead{marc.avila@zarm.uni-bremen.de}
\cortext[cor1]{Corresponding author}
\address[1]{University of Bremen, Center of Applied Space Technology and Microgravity (ZARM), 28359 Bremen, Germany}
\address[2]{Tianjin University, Center for Applied Mathematics, Tianjin 300072, China}
\address[3]{Institute of Science and Technology (IST), Am Campus 1, 3400 Klosterneuburg, Austria}


\begin{abstract}
Phase-field methods have long been used to model the flow of immiscible fluids. Their ability to naturally capture interface topological changes is widely recognized, but their accuracy in simulating flows of real fluids in practical geometries is not established. We here quantitatively investigate the convergence of the phase-field method to the sharp-interface limit with simulations of two-phase pipe flow. We focus on core-annular flows, in which a highly viscous fluid is lubricated by a less viscous fluid, and validate our simulations with an analytic laminar solution, a formal linear stability analysis and also in the fully nonlinear regime. We demonstrate the ability of the phase-field method to accurately deal with {non-rectangular} geometry, strong advection, unsteady fluctuations and large viscosity contrast. We argue that phase-field methods are very promising for quantitatively studying moderately turbulent flows, especially at high concentrations of the disperse phase.
\end{abstract}

\begin{keyword}
phase-field method \sep pipe flow \sep hydrodynamic stability
\end{keyword}

\end{frontmatter}


\section{Introduction}
\label{sec:introduction}

Many numerical methods have been developed to deal with the motion of interfaces between immiscible fluids \cite{Tryggvason2009}. 
These methods can be divided into two broad classes: interface-tracking and interface-capturing methods 
\cite{Cristini2004}. In interface-tracking (or sharp-interface) approaches, the governing (Navier--Stokes) equations
are solved separately for different phases and the coupling occurs through stress and velocity boundary conditions 
at interfaces. This requires explicit tracking and meshing of the interfaces \cite{Tryggvason2001}. 
Because of the computational cost incurred in re-meshing the interface, volume-of-fluid (VoF) \cite{Hirt1981} 
and level-set (LS) methods \cite{Sussman1994} were developed to `capture' the 
evolution of the interface. In these methods, a single set of the Navier--Stokes 
equation for the whole domain is solved together with an advection equation of a scalar field (the volume fraction 
in VoF and a distance function in LS). The surface-tension force appears then in the Navier--Stokes equations
as a volume-force applied in a narrow region near the interface, based on the Young-Laplace formula. These methods yield good approximations of the sharp discontinuities across interfaces, but require the calculation of the surface-normal vector and can lead to difficulties in resolving large interface curvatures. Furthermore, the cost of computing the curvature of the interface rises linearly with the interfacial area. Thus, in problems with large interfacial areas and deformations, such as in non-dilute turbulent dispersions, the additional computational cost of these methods is substantial \cite{Mirjalili2018}.

\subsection{Cahn--Hilliard--Navier--Stokes equations} \label{sec:CH-method}

An alternative interface-capturing approach to computing multiphase flows is the diffusive-interface (or phase-field) 
method \cite{Anderson1998}. Here the sharp changes occurring across the interface are 
smoothed over a finite-thickness mixing layer of width $\epsilon$ separating the phases. Across this mixing layer, 
the phase-field variable $C$, which describes the composition of the mixture, transitions 
from the constant value in one phase to that in the other.  In our formulation $-0.5\leqslant C \leqslant 0.5$ is chosen, so that $C=-0.5$ and $0.5$ correspond to fluid 1 and 2, respectively. The density and viscosity of the fluids are typically defined as 
\begin{equation}\label{eq:rho_nu}
\rho(C) =\frac{\rho_1+\rho_2}{2} + (\rho_2-\rho_1)C,~~~
\mu(C)  =\frac{\mu_1+\mu_2}{2} + (\mu_2-\mu_1)C,
\end{equation}
for simplicity. The interfacial dynamics is modeled accounting for the free energy of the two-fluid system
\begin{equation}\label{equ:free_energy}
\mathcal{F}=\int_\text{V}\dfrac{\alpha}{2}\left(\nabla C\right)^2 + \beta\Psi(C)\,d\text{V},
\end{equation}
where the material parameters  $\alpha$ and $\beta$ determine the contribution of the gradient energy and bulk energy, respectively. The double-well potential $\Psi(C)=(C+0.5)^2(C-0.5)^2$ models the immiscibility of the two 
fluids. Equilibrium interface profiles are those that minimize the free energy \eqref{equ:free_energy}. The interface thickness
$\epsilon$ and surface tension $\sigma$ of the system emerge from the competition between the gradient term, which tends to broaden the interface, and the bulk term, which favors phase-separation. In the isothermal case,  $\sigma=\sqrt{\alpha\beta/18}$ and the interface thickness { $\epsilon$} is typically defined as the region for which $-0.45\leqslant C \leqslant 0.45$. With this choice $\epsilon = 4.164\sqrt{\alpha/\beta}$ and the so-defined interface contains about 98.5\% of the surface tension stress \cite{Jacqmin1999}. The chemical potential of the system
\begin{equation}\label{chemical_potential}
\Phi=\dfrac{\delta \mathcal{F}}{\delta C}=\beta\Psi'(C)-\alpha\nabla^2C, 
\end{equation}
is the rate of change of the free energy $\mathcal{F}$ with respect to $C$, and so equilibrium interface profiles satisfy $\Phi$=0 \cite{Jacqmin1999}. For a planar interface, $C$ has a hyperbolic tangent profile \cite{Jacqmin1999}. 
The temporal evolution of $C$ is governed by the Cahn--Hilliard equation \cite{Cahn1958} 
\begin{equation}\label{equ:Cahn-Hilliard}
\frac{\partial C}{\partial t}+\boldsymbol{u}\cdot\nabla C = \kappa\nabla^2\Phi,
\end{equation}
where $\kappa$ is the mobility of the chemical potential. The Cahn--Hilliard equation is solved together with the Navier--Stokes equations (CHNS), 
\begin{align}
\label{N-S}\rho\left(\frac{\partial \boldsymbol{u}}{\partial t}+\boldsymbol{u}\cdot\nabla\boldsymbol{u}\right) &= -\nabla \tilde p + \nabla\cdot\left(\mu\left(\nabla \boldsymbol{u}+(\nabla\boldsymbol{u})^T\right)\right)-C\nabla\Phi + \rho\boldsymbol g + \boldsymbol f,\\
\label{incompressibility} \nabla\cdot\boldsymbol u &=0,
\end{align}
where $\boldsymbol f$ represents external body forces and $-C\nabla\Phi$ is the motion-causing component of the surface-tension. Its potential component has been absorbed into the generalized pressure $\tilde p=p - C\Phi + \beta\Psi - \alpha/2|\nabla C|^2$, where $p$ is the  true fluid pressure \cite{Jacqmin1999}. Note that $C$ also enters the Navier--Stokes equations through the variable density and viscosity of the mixture \eqref{eq:rho_nu}.

The CHNS 
have been long used to simulate binary fluid flows \cite{Anderson1998} and more recently to investigate turbulent multiphase flows \cite{Scarbolo2015,Ahmadi2016}. Their ability to deal with topological changes and their thermodynamic consistency make them an appealing alternative to other methods. It has been shown that the classical stress balance at fluid-fluid interfaces (employed in sharp-interface methods) is recovered from the CHNS in the limit of vanishingly small interface width $\epsilon\rightarrow0$, see  \cite{Anderson1998}. Note that this is not the case for other interface-capturing methods, such as volume-of-fluid methods, which may explain their difficulties in producing grid-converged results in some specific problems \cite{Klostermann2013}. 

\subsection{Interface thickness and mobility}

For immiscible fluids, $\epsilon$ is a few nanometers wide (e.g.~as measured for water and n-alkyl \cite{Mitrinovic2000}) 
and the mobility of the chemical potential, though not directly measurable in experiments, was estimated to be  $\kappa\sim 10^{-17}\text{m}^3\text{kg}^{-1}\text{s}$ \cite{Jacqmin1999}. Such values are far beyond the ability of state-of-the-art numerical simulations for  problems outside the domain of nanofluidics. In practice, the interface thickness has to be resolved with a few grid points, and this raises the question of how to  choose the value of the mobility $\kappa$ consistently. Starting with Jacqmin \cite{Jacqmin1999}, who argued that the mobility should scale as $\kappa\propto \epsilon ^\delta$, with $1\leqslant\delta<2$, several scalings have been proposed and tried, see e.g.~\cite{Khatavkar2006,Yue2010}. Hence, despite the solid theoretical footing and promise of diffuse-interface methods, their predictive power has remained poor because $\kappa$ has been essentially used as a free numerical parameter in previous studies. { An improper choice for the mobility parameter (i.e.\ not consistent with the sharp-interface limit) can lead to a numerical inaccuracy, which hinders the application of the phase-field method to two-phase flows in engineering.} 

Magaletti et al.~\cite{Magaletti2014} resolved this controversy recently. They carried out an asymptotic analysis of the CHNS system for $\epsilon\to 0$ and showed that only the scaling $\kappa\propto\epsilon^2$ can correctly match the inner (interface) and outer (bulk) dynamics. This is necessary to recover the correct surface tension force. Physically, the mobility determines the time scale of the diffusion of the chemical potential and cannot be neither too large nor too small (for a given $\epsilon$). If the mobility is too small, then the inner dynamics responds too slowly to advection by the fluid velocity field and the surface-tension force is incorrect. In contrast, if the mobility is too large, then the interface is hardly deformed by the velocity field. In other words, the Cahn--Hilliard equation has to generate the correct $C$ (and $\Phi$) profile at the right rate, i.e.~following the change in the outer flow, and this requires a consistent value of the mobility. Magaletti et al.~\cite{Magaletti2014} further verified their theoretical derivation with simulations of capillary waves and drop coalescence in two-dimensional Cartesian geometry for matched density and viscosity. In addition, they pointed out that the inner dynamics of the interface, which is much faster than the outer time scale of the fluid velocity field, must be resolved to correctly recover the interfacial physics. Although this can potentially pose a more stringent restriction on the time-step size than numerical stability, it has been scarcely addressed in the literature \cite{Magaletti2014}. 

\subsection{Core-annular flows in pipes} \label{sec:core-annular}

\begin{figure}
\centering                                                                                                                                                             
		\begin{tikzpicture}
		\node [anchor=west] (start) at (4.2,4.2) {\Large  };
		\begin{scope}
	    \node[anchor=south west,inner sep=0] (image) at (0,0) 			{\includegraphics[width=0.6\textwidth]{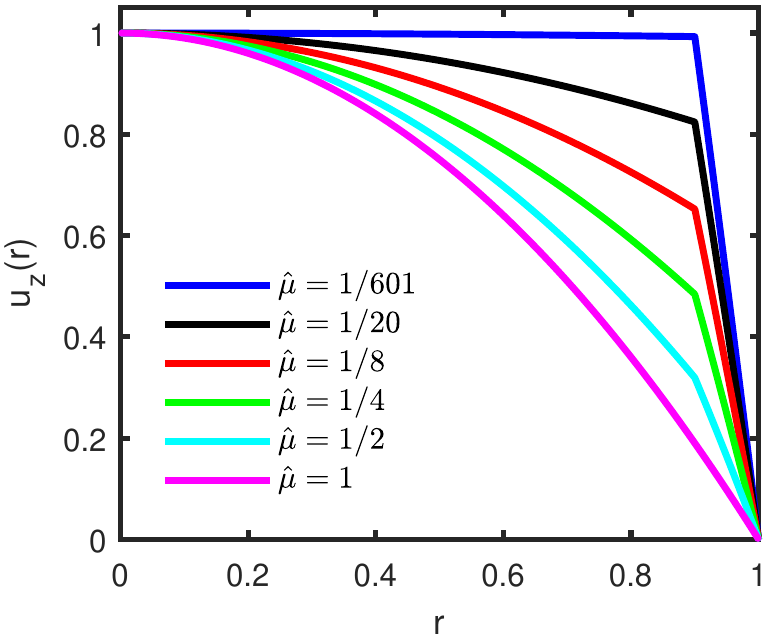}};
		\draw [-stealth, line width=1.5pt, black] (start) -- ++(2,2);
		\end{scope}
\end{tikzpicture}%
\caption[base flow]{\label{fig:basic_flow} 
Axial velocity profile of laminar core-annular flow $u_z(r)$ of density-matched fluids in the sharp-interface limit for $\eta=R_I/R=0.9$, where $R$ is the pipe radius and $R_I$ the position of the interface. The radial position $r$ is normalized with  $R$, whereas the axial velocity is normalized with its centerline value $u_z(0)=U$. The arrow indicates the direction of increasing viscosity ratio of the annular to the core fluid, as indicated in the legend.
}     
\end{figure}

The purpose of this paper is to investigate the numerical convergence of the phase-field model to the sharp interface limit as $\epsilon\to 0$, thereby assessing the robustness of  the scaling  $\kappa\propto \epsilon^2$, in a situation of engineering interest, i.e.\ with real fluids, practical geometries and flow rates. We focus on the study of core-annular flows (CAF), which are a particular set of two-phase pipe flow regimes where an inner fluid, the ``core", is surrounded by an outer fluid, the ``annulus". These flows are of interest because they allow the lubricated transport of viscous fluids. In the core-annular configuration, the viscous fluid (normally a heavy oil) tends towards the center of the pipe while the less viscous one (usually water) migrates to the high-shear region close to the walls. This structure results in a large reduction in the friction losses when compared with the flow of the single-phase viscous fluid. { The interested reader is referred to Joseph \emph{et al}.~\cite{Joseph1997} for a comprehensive review of CAF, and to Govindarajan and Sahu \cite{Govindarajan2014} for a more recent review focusing on miscible and immiscible viscosity-stratified flows.} By assuming an infinite or axially periodic pipe, the Navier--Stokes equations with stress and velocity boundary conditions at the interface between the two fluids admit an analytic steady solution termed laminar core-annular flow, for which the interface between the core and annular fluids is parallel to the pipe wall. Figure~\ref{fig:basic_flow} shows the velocity profiles of laminar CAF for several viscosity ratios and matched densities. When both fluids have the same viscosity, the classic parabolic Poiseuille profile is recovered. As the viscosity ratio increases, the velocity profile becomes steeper in the annular region, whereas in the core the velocity tends to become uniform, i.e.~the core exhibits solid-like behavior.

In this paper, we present a comprehensive investigation of the convergence of the CHNS to the sharp-interface limit for CAF with axially periodic boundary conditions. First, the dependence of the convergence rate on the viscosity contrast is investigated for laminar CAF.  Second, we examine the convergence of the linear instability of the laminar CAF as a function of the mobility and time-step size. Finally, we compute flow patterns in the fully nonlinear regime and compare them quantitatively to simulations with the VoF method. Overall, our results highlight the ability of the phase-field method to produce accurate results in practical flow situations.

\section{Problem specification and methods}

We consider upward CAF where the core fluid is oil (subindex o) and the annular fluid is water (subindex w). Here, the flow is driven upward against gravity by imposing a negative average pressure gradient in the axial direction,  $\mathrm{d}\overline{p}/\mathrm{d}z<0$. In this case, the centerline velocity of laminar CAF reads 
\begin{equation}\label{eq:U}
U = (f-\rho_o g)\dfrac{R_I^2}{4 \mu_o} + (f - \rho_w g)\dfrac{R^2-R_I^2}{4 \mu_w} - (\rho_o - \rho_w)g\dfrac{R_I^2}{2\mu_w}\log \dfrac{R}{R_I}, 
\end{equation}
where $f=-\mathrm{d}\overline{p}/\mathrm{d}z>0$, $R$ is the radius of the pipe, $R_I$ the interface position and $\mu_o$ ($\mu_w$) and $\rho_o$  ($\rho_w$) are the dynamic viscosity and the density of oil (water). In this paper, all variables and parameters are rendered dimensionless by using the following transformations
\begin{align}\label{normalization}
\boldsymbol u^*=\dfrac{\boldsymbol u}{U}, \quad \boldsymbol x^*=\dfrac{\boldsymbol x}{R}, & \quad t^*=\dfrac{U\,t}{R}, \quad p^*=\dfrac{R\,p}{\mu_o\,U},   \nonumber \\ 
\rho^*=\dfrac{\rho}{\rho_o},\quad \mu^*= & \dfrac{\mu}{\mu_o},\quad f^*=\dfrac{f R}{\rho_o U^2}.
\end{align}
Hereafter, only dimensionless quantities are used and the superscript $^*$ is dropped to simplify the notation. 

\subsection{Governing equations and parameters}

The dimensionless CHNS equations for upward CAF read
\begin{align}
\label{dimlessN-s}\rho Re \left(\dfrac{\partial\boldsymbol u}{\partial t} +  \boldsymbol u \cdot \nabla \boldsymbol u \right)  &= - \nabla \tilde{p} + \nabla \cdot  \boldsymbol T -  \frac{\sqrt{18}}{CaCn}C\nabla \Phi  + Re \left(f-\frac{\rho}{Fr} \right)\boldsymbol e_z,\\
\nabla \cdot \boldsymbol u &= 0,\\
\dfrac{\partial C}{\partial t} + \boldsymbol u\cdot \nabla C &= \frac{1}{Pe} \nabla^2 \left( \Psi'\left( C \right) - Cn^2\nabla^2C \right), \label{equ:CH_dimless}
\end{align}
where  $ \boldsymbol T=\mu( \nabla \boldsymbol u + \nabla \boldsymbol u^T)$ is the viscous stress tensor. The viscosity ratio and the density ratio
\begin{equation}\label{material_params}
\hat\rho=\dfrac{\rho_w}{\rho_o},\qquad \hat \mu=\dfrac{\mu_w}{\mu_o}, 
\end{equation}
enter the equations through the dimensionless viscosity $\mu$ and density $\rho$ and are solely set by the choice of fluids, whereas the Reynolds, capillary and Froude numbers
\begin{equation}\label{dyn_params}
Re=\dfrac{\rho_oU R}{\mu_o},\qquad  Ca=\dfrac{\mu_o U}{\sigma},\qquad Fr=\dfrac{U^2}{R g},
\end{equation}
depend also on the fluid velocity. {  The surface tension effects in the Cahn-Hilliard approach are covered by three dimensionless numbers, namely the capillary number, the Cahn number and the Peclet number. The latter two, which are defined as}
\begin{equation}\label{chns_params}
Cn=\dfrac{\sqrt{\alpha/\beta}}{R},\qquad
Pe=\dfrac{U R}{\kappa\beta},
\end{equation}
are the dimensionless interface width and the dimensionless inverse of the mobility, respectively, and determine how the sharp-interface limit is approached. { In dimensionless form, the relationship} $\kappa\propto\epsilon^2$ derived by Magaletti \emph{et al}.~\cite{Magaletti2014} turns into  
\begin{equation}\label{equ:Pen_relation}
Pe\propto Cn^{-1}.
\end{equation}

\subsection{Boundary conditions and hold-up ratio}

Bai \emph{et al}.~\cite{Bai1992} investigated experimentally CAF of very viscous oils lubricated with water. In their experiments, the input volume flow rates of oil ($\dot V_o$) and water ($\dot V_w$) were set independently. This setup could be modeled numerically by using inlet boundary conditions in a pipe, but Li and Renardy~\cite{Li1999} pointed out that the slow development of the ensuing flow patterns in the streamwise direction makes such simulations infeasible. Instead, they performed simulations with axially periodic boundary conditions, as done in this paper. Note, however, that prescribing both input volume flow rates is not possible in axially periodic pipes, where the two adjustable parameters are the ratio of oil-to-water volumes,
\begin{equation}
 \hat V=\dfrac{V_o}{V_w}, \label{V_ratio}
\end{equation}
and the pressure gradient used to drive the flow. The former is fixed by the initial condition, whereas the latter is imposed with a volume force in the right-hand-side of the Navier--Stokes equation ($f$ in eq.~\ref{dimlessN-s}). 

For laminar CAF, it is  easy to relate the parameters in numerical simulations of axially periodic pipes to those used in experiments {because of the existence of the following analytical solution} 
\begin{equation}\label{equ:basic_flow}
u_z(r)=
\begin{cases}
\left(\dfrac{1}{\eta^2} - \dfrac{r^2}{\eta^2} - 2(K_f-1)\text{log}(r)  \right)\dfrac{1}{A} &\text{if } 0<r\leqslant \eta,\\
\qquad \qquad 1 - \hat{\mu}\dfrac{r^2}{\eta^2}\dfrac{K_f}{A} &\text{if } \eta<r\leqslant 1,
\end{cases}
\end{equation}
where $\eta$ is the dimensionless position of the interface, related to the volume ratio as $\hat V=\eta^2/(1-\eta^2)$, $K_f=(Fr\,f-1)/(Fr\,f - \hat\rho)$ is the driving force ratio and $A=\hat{\mu}K_f + 1/\eta^2 - 1 + 2(K_f-1)\log(1/\eta)$. {By contrast, an analytical solution is not available for the nonlinear regimes, which makes the comparison between numerical simulations and experiments more challenging.} The hold-up ratio
\begin{equation}\label{eq:holdup}
 h=\dfrac{\hat{\dot{V}}}{\hat{V}},
\end{equation} 
where $\hat{\dot{V}}=\dot{V}_{\text o}/\dot{V}_{\text w}$, is frequently used to characterize two-phase pipe flows~\cite{Bai1992}. It is a measure of the relative velocities of both fluids, and in case of a perfectly mixed flow it should have a value of one. In CAF, it is expected that the liquid in contact with the pipe wall is held back and therefore the holdup ratio should be larger than one. 
To enable comparison with experiment, Li and Renardy~\cite{Li1999} extracted the volume ratio from the experimentally measured hold-up ratio, see eq.~\eqref{eq:holdup}. Then, by using eq.~\eqref{equ:basic_flow}, thereby assuming laminar CAF, they computed the pressure gradient yielding the experimentally imposed oil volume flow rate $\dot{V}_o$. Note, however, that with this procedure, the water volume flow rate $\dot{V}_w$ is fixed indirectly and is only correct {in the laminar CAF regime (i.e.\ in the absence of interface waves).} 

At the pipe wall, no-slip boundary conditions were applied for the velocity field and no-flux boundary conditions  for the phase-field variable and chemical potential. The latter corresponds to 90$^\circ$ contact angle at the pipe wall \cite{Jacqmin1999}. Note however that this choice is irrelevant because in our simulations the core fluid never touched the wall. 

\subsection{Numerical method}\label{sec:method}

We used the strategy of \cite{Dong2012} to solve the Cahn-Hilliard equation and to treat the variable coefficient of the Laplacian terms in the Navier-Stokes equations. { For the time-integration scheme (Crank--Nicolson) and the enforcement of the incompressibility condition \eqref{incompressibility} with the influence matrix method, we followed the approach of the open-source pipe flow solver \textit{openpipeflow} \cite{openpipeflow}.  
For further details on the time-integration scheme, we refer the reader to \cite{Guseva2015}.} We solved the CHNS in cylindrical coordinates $(r,\theta,z)$ and used 7-point finite-difference stencils to discretize the radial direction, whereas the Fourier--Galerkin spectral method was used in the periodic axial and azimuthal directions. Variables ($\boldsymbol u$, $p$ and $C$) were  written as
\begin{equation}\label{equ:Fourier_expansion}
A(r,\theta,z,t)=\sum_{k=-K}^{K}\sum_{m=-M}^{M}\hat{A}_{k,m}(r,t)e^{(ik_0 kz+im\theta)},
\end{equation}
where $k_0k$ and $m$ are the wave numbers of the modes in the axial and  the azimuthal directions, respectively. The pipe length is $L_z=2\pi/k_0$, and $\hat{A}_{k,m}$ is the complex Fourier coefficient of mode $(k,m)$. 

In the radial direction, the interfacial region ($-0.45<C<0.45$) was discretized with $8$ points, corresponding to $\Delta r \approx 0.5 Cn$. This criterion serves to estimate  the required resolution for simulations of the CHNS and poses a stark restriction. However, in problems where the interface position is fixed, it can be alleviated by using a non-uniform grid. For the tests presented here we used various grids, all of them presenting a clustering of the radial points close to the wall and near the interface (since its position varied little through the simulation). This  rendered a manageable number of grid points in the radial direction, even for very small values of $Cn$, and ensured that spatial discretization errors were negligible. This approach enabled extensive tests of the convergence of the method as a function of the viscosity ratio, $Cn$, $Pe$ and time step, which was one of the main purposes of this work.

The evaluation of the nonlinear terms was performed using the pseudo-spectral technique (with the $3/2$-rule for de-aliasing), which utilizes the fast Fourier transform (FFT) to convert data between physical and spectral spaces. Solving the CH-equation and the full treatment of the viscous term, due to the variable density and viscosity, implies that this method requires roughly 2.5 times more operations for each evaluation of the non-linear terms than a similar single-phase code using the same discretization and time-stepping scheme. We used the MPI-OpenMP hybrid parallelization strategy of \cite{Shi2015}, which can efficiently utilize $\mathcal{O}(10^4)$ processors for high-resolution simulations \cite{rampp2015nscouette}.


\section{Laminar core-annular flow in the phase-field formulation}\label{sec:basic_flow}

We first examined the convergence of the laminar CAF solution to the CHNS equations to the analytic sharp-interface solution \eqref{equ:basic_flow}. For simplicity, we focused on matched density fluids, for which the volume force required to drive the flow in the right-hand-side of the Navier--Stokes equation~\eqref{dimlessN-s} reads
\begin{equation}\label{equ:analytical_forcing}
  f=\frac{1}{Re} \dfrac{4 \hat{\mu}}{1-\eta^2+\eta^2\hat\mu}.
\end{equation}
In the phase-field formulation there is, to our knowledge, no analytical expression for laminar core-annular flow, so we computed it numerically. Under base flow conditions (radial dependence only, steady) the left-hand-side of the Cahn--Hilliard equation is zero and the basic flow is independent of $Pe$. Moreover, in this case the surface tension force does not cause fluid motion and is thus zero in the potential formulation used here. Hence the Cahn--Hilliard equation decouples from the Navier--Stokes equations and the convergence of the base flow depends only on the interface thickness and the viscosity ratio. { The system was solved in one dimension (radial coordinate), in order to obtain the laminar CAF solution without interfacial waves.}

\begin{figure}
\centering                                                                                                                                                             
\begin{tikzpicture}
		\node [anchor=west] (start) at (0,1.5) {\Large  };;
		\begin{scope}
	    \node[anchor=south west,inner sep=0] (image) at (0,0) 			{\includegraphics[width=0.55\textwidth]{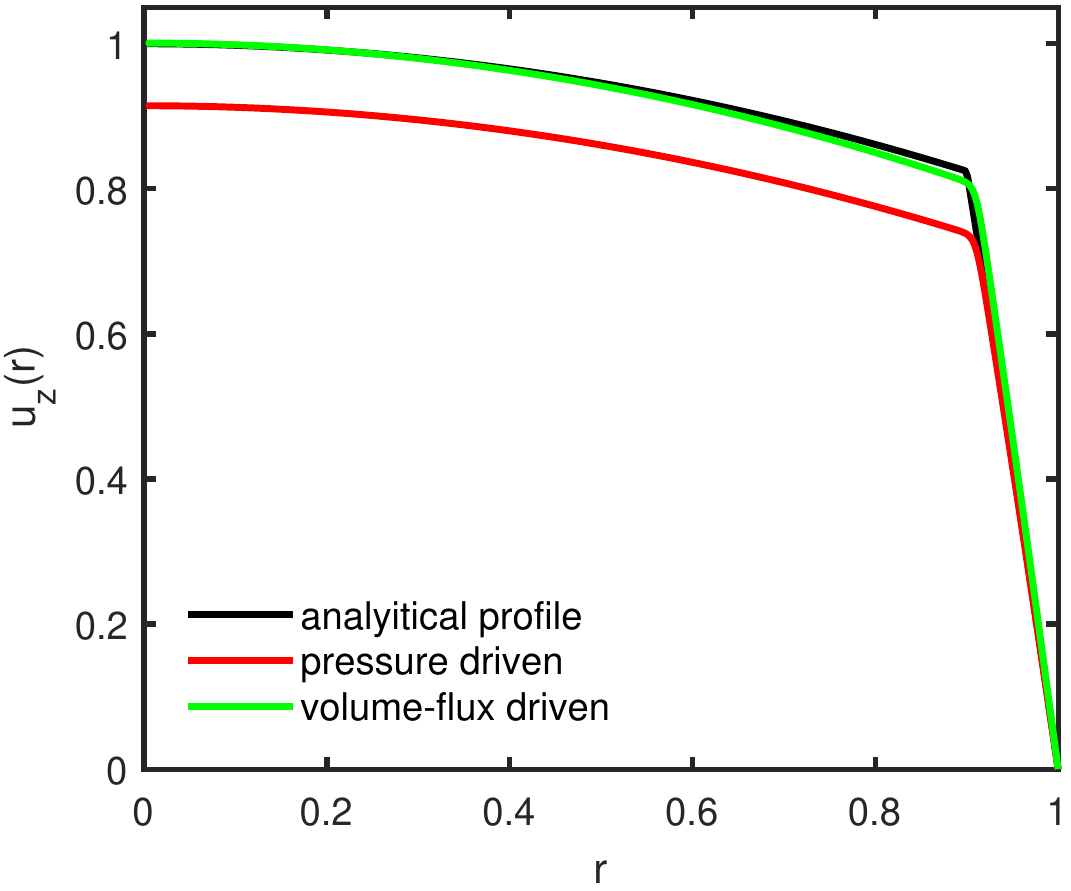}};
	    \node[anchor=south west,inner sep=0] (image) at (8,0.4) 			{\includegraphics[width=0.35\textwidth]{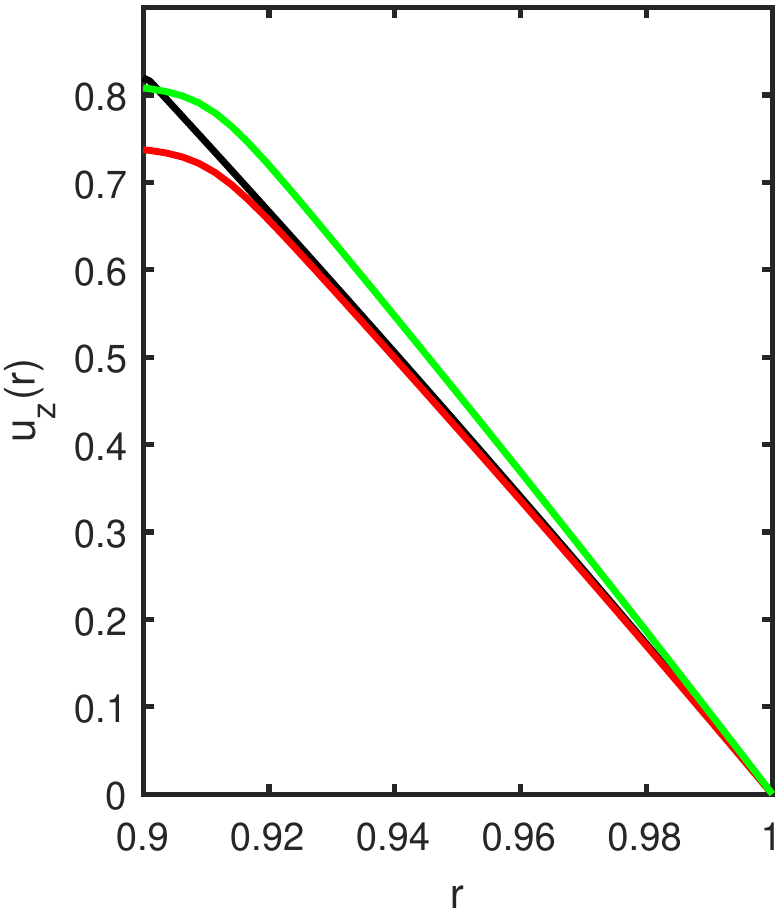}};
	    \draw [dashed, color=red] (7.1,5.5) -- (8.85,5.95);
	    \draw [dashed, color=red] (7.1,0.848) -- (8.85,1.13);
	    \draw [dashed, color=red] (6.805,0.849) -- (7.46,0.849)-- (7.46,5.46)-- (6.805,5.46)--(6.805,0.849);
		\node [anchor=west] (g) at (3.8,6.8) {(a) };
		\node [anchor=west] (g) at (10.5,6.8) {(b) };
		\end{scope}
\end{tikzpicture}%
\caption[base flow]{\label{fig:CH_sol_laminarflow} 
(a) Laminar CAF with $\eta=0.9$ (volume ratio $\hat V=4.26$) and $\hat{\mu} = 1/20$. Red and green lines correspond to pressure-driven and volume-flux driven flow solutions of the CHNS with $Cn = 0.005$. The black line is the reference analytical solution~\eqref{equ:basic_flow}  in the sharp-interface limit. (b) close-up of the profiles near the wall. (For interpretation of the references to colour in this figure legend, the reader is referred to the web version of this article.)}     
\end{figure}
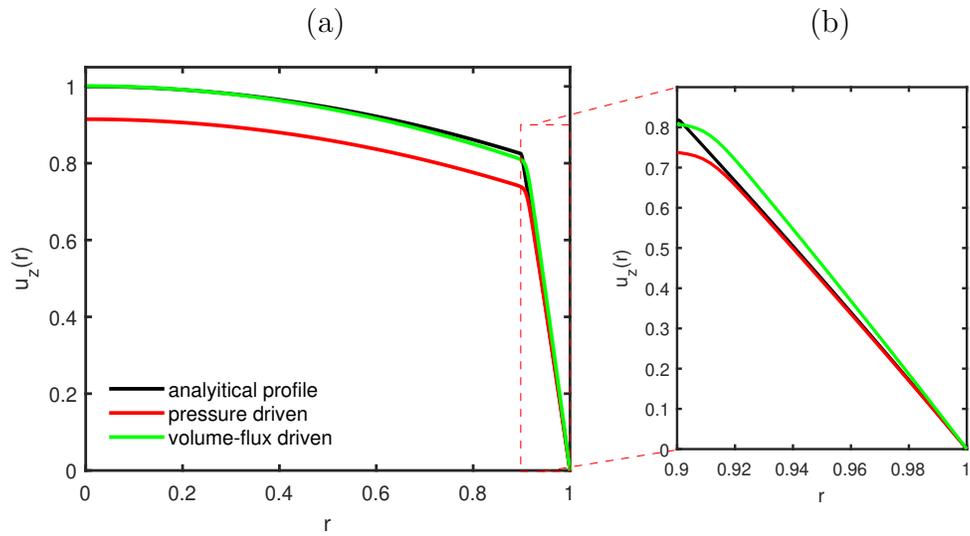

After initial transients, our simulations converged toward the velocity profile shown as a red line in figure~\ref{fig:CH_sol_laminarflow}. This agrees excellently with the analytical solution (black line) on the annular side, but is significantly slower in the core region. In particular, the centerline velocity does not reach the value $u_z(0)=1$ expected for the analytic solution. A better numerical approximation to the analytical solution in the core is achieved when the fluid motion is driven by imposing a constant (total) volume flux along the pipe (green line). In this case, the driving volume force $f$ is adjusted every time-step so that the total volume flux is identical to that of the analytical solution, namely
\begin{equation}
\dot{V}=\dfrac{\pi}{2}\,\dfrac{1 - \eta^4(1 - \hat{\mu})}{1 - \eta^2(1-\hat{\mu})}.
\end{equation}

\begin{figure}[!ht]
        \centering
        \begin{subfigure}[b]{0.475\textwidth}
            \centering
             \caption%
            {{\small }} 
            \includegraphics[width=\textwidth]{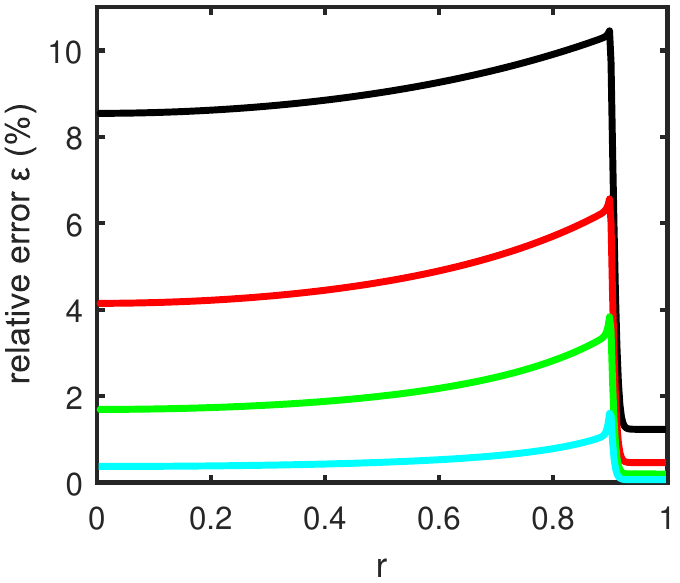}
               \label{fig:baseflow_error_f}
        \end{subfigure}
        \hfill
        \begin{subfigure}[b]{0.475\textwidth}  
            \centering 
            \caption[]%
            {{\small }}   
            \includegraphics[width=\textwidth]{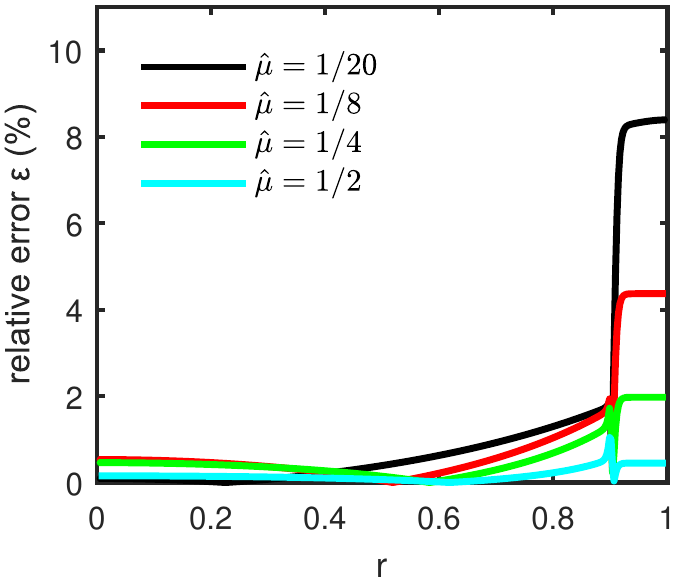}
           \label{fig:baseflow_error_Q}
        \end{subfigure}
        \vskip\baselineskip
        \begin{subfigure}[b]{0.475\textwidth}   
            \centering 
            \caption[]%
            {{\small }}  
            \includegraphics[width=\textwidth]{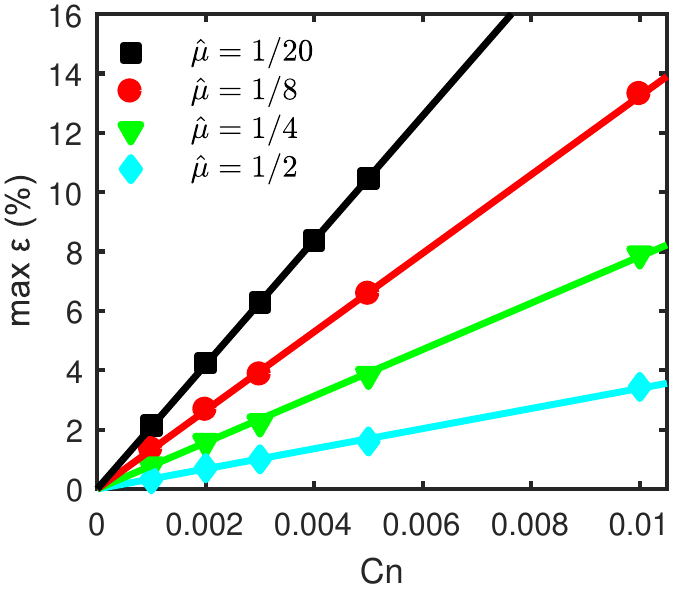}  
            \label{fig:baseflow_conv_f}
        \end{subfigure}
        \quad
        \begin{subfigure}[b]{0.475\textwidth}   
            \centering 
            \caption[]%
            {{\small }} 
            \includegraphics[width=\textwidth]{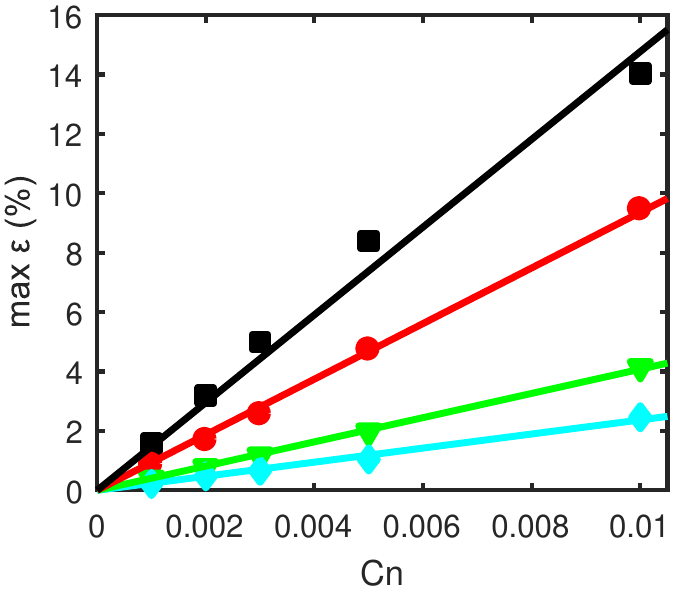}
            \label{fig:baseflow_conv_Q}
        \end{subfigure}
        \caption[base flow convergence]{\label{fig:basic_flow_convergence} 
Relative error of the numerically computed laminar CAF with respect to the sharp-interface limit solution for $\eta = 0.9$ and viscosity ratio $\hat\mu$ as indicated in the legends. The left (right) panels correspond to pressure-driven (volume-flux driven) flow. 
(a)--(b) Radial distribution of the error for $Cn=0.005$.  (c)--(d) Maximum error as a function of  $Cn$. The solid lines are linear fits to the data. }
    \end{figure}
    
{ The relative error of the computed laminar solution with respect to the analytical one is shown in figure~\ref{fig:basic_flow_convergence}a--b as a function of the radial coordinate for pressure-driven and volume-flux driven flows, respectively. It is interesting to note that despite the difference in the profiles apparent in figure~\ref{fig:CH_sol_laminarflow}, the maximum error is of similar magnitude and occurs near the interface between the two fluids in both cases. Figure~\ref{fig:basic_flow_convergence}a confirms that if the pressure gradient of the analytical solution is imposed, the error is concentrated in the core side and decreases toward the wall. In fact, because of axial momentum conservation, with pressure driving the shear stress at the wall should be equal to the analytical one. However, in the Cahn-Hilliard method, there is a slight shift in $C$ in the bulk phases with respect to the expected values, $C=\pm0.5$. This shift is proportional to $Cn$ \cite{Yue2007}, and causes an error in the calculation of the viscosity in eq.~\eqref{eq:rho_nu}. Hence, even in the pressure-driven case, there is a discrepancy with the analytical profile in the vicinity of the wall. For the volume-flux driven case, the error near the wall is much larger. In particular, imposing the same volume flux as for the analytical solution results in larger shear stress at the wall (see figure ~\ref{fig:CH_sol_laminarflow}b) and thus larger pressure gradient to drive the flow.}

\begin{figure}[!ht]
\centering	
\includegraphics[width=0.75\linewidth]{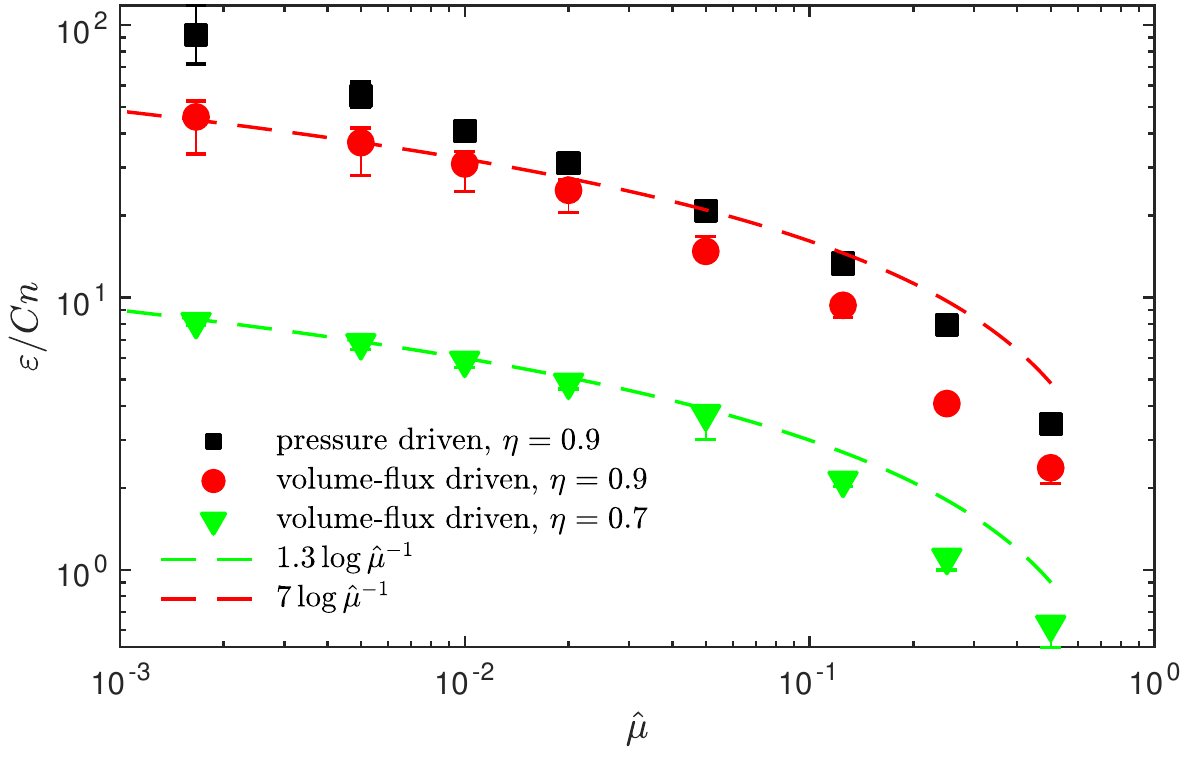}

\caption[base flow convergence]{\label{fig:convergence_rate_vs_muratio} 
Dependence of the maximum error (normalized with the Cahn number) on the viscosity ratio of the laminar CAF. { The data shown as \mysquare{black} and \mycircle{red} are the proportionality constants of the solid lines in figure~\ref{fig:basic_flow_convergence}c--d, respectively, whereas the error bars quantify the difference between the solid lines (fits) and the data points. Note that additional computations were performed for even lower values of the viscosity ratio ($\hat\mu=1/50$, $1/100$, $1/200$ and $1/601$) and are also included here.   The data shown as \mydowntriangle{green} are from another case with lower volume ratio, $\hat V=0.96$. The dashed lines show the approximation to the error given by eq~\eqref{eq:err_lam}, which is only valid  for small $\hat\mu$.}}
\end{figure}

The maximum error with respect to the analytical solution is shown in figures~\ref{fig:basic_flow_convergence}c--d as a function of $Cn$ for several $\hat\mu$. Regardless of the driving and the viscosity ratio used, the computed base flow converges linearly to the analytical solution as $Cn$ decreases. This is consistent with previous studies in which the phase-field method was postulated to be first-order accurate with respect to $Cn$~\citep{Jacqmin1999,Magaletti2014}. { The effect of the viscosity ratio $\hat\mu$ on the error was investigated quantitatively by determining the proportionality constants of the lines in figure~\ref{fig:basic_flow_convergence}c--d. These are shown as squares and circles, respectively, in figure~\ref{fig:convergence_rate_vs_muratio}. Clearly, as $\hat\mu$ decreases toward zero, corresponding to very viscous oils in the core, the error increases monotonously. We also computed another case with $\eta=0.7$  (volume ratio $\hat V=0.96$) for constant volume-flux driving. The corresponding error is shown as triangles and is substantially smaller than for $\eta=0.9$ ($\hat V=4.26$). This is as expected because the shift in $C$ in the bulk phases becomes worse, the more that $\hat V$ departs from unity \cite{Yue2007}.}

{ In order to qualitatively understand the effect of viscosity ratio on the error shown in figure~\ref{fig:convergence_rate_vs_muratio}, we developed a simple model. We assumed that the equilibrium profile of $C$ approaches a hyperbolic tangent as $Cn\rightarrow 0$ \cite{Jacqmin1999}. Then by using this profile in eq.~\eqref{eq:rho_nu} and integrating the one-dimensional steady Navier--Stokes equation, an approximation of the velocity profile was obtained. This model profile also converges to the analytic laminar CAF as $Cn\rightarrow 0$. In particular, it can be shown that at small $\hat\mu$ and small $Cn$, the error of the model velocity profile is 
\begin{equation}\label{eq:err_lam}
 \varepsilon =\mathcal{O}(Cn\log\hat{\mu}^{-1}).
\end{equation}
The dashed lines in figure~\ref{fig:convergence_rate_vs_muratio} shows that this approximation is indeed very good for low viscosity ratio and constant volume-flux driving. For pressure-driven flow there is more scatter in our data and the agreement is less good (not shown here). Note also that because the assumed $\tanh$-profile for $C$ in the model does not exhibit a shift in the bulk phases, eq.~\eqref{eq:err_lam} does not capture the dependence of the error on  the volume ratio $\hat V$.}  

{ }

\section{Linear stability analysis of core-annular flow}\label{sec:linstbanl}

Hu and Joseph~\cite{Hu1989} studied the linear stability of density-matched laminar CAF and showed that this is stable only in certain parameter regimes. Furthermore, they did an energy analysis for $Ca=0.5$, $\hat \mu = 0.1$ and $\eta=\{0.7,0.8\}$ and elucidated the contribution of all the terms in the energy equation, thereby classifying the underlying instabilities according to the dominant physical mechanisms. At low $Re$, capillary instability dominates, whereas as $Re\gtrsim50$ the interfacial-friction due to the difference in viscosity becomes predominant. For $Re\gtrsim200$ the production of energy in the bulk of the less viscous (annular) fluid exceeds its dissipation and the instability is predominantly inertial.

\begin{table}
\centering
\begin{tabular}{|c|c|c|c|c|c|c|c|}
\hline
 Test case & $Re$ & $Ca$ &  $\hat\mu$ & $\hat\rho$ 
            & $\eta$ & $\hat V$\\ 
\hline
CAF1 & 37.78 & $\infty$  & 1/2 & 1  & 0.7  & 0.9608\\ 
\hline
CAF2 & 500 & 0.5  & 1/20 & 1  & 0.9  &4.2632\\
\hline
\end{tabular}
\caption{\label{tab:test_cases} Parameters of the two cases considered for the linear stability analysis, following Hu \& Joseph~\cite{Hu1989}. In both cases, the most unstable disturbance is axisymmetric ($m=0$), evolving into an interfacial-friction (CAF1) or inertial (CAF2) instability.}
\end{table}

We computed the linear-stability of laminar CAF for two specific cases discussed by Hu and Joseph~\cite{Hu1989} and specified in table~\ref{tab:test_cases}. The flow was driven by a constant volume flux and { simulations were run with a small number of Fourier modes in the axial and azimuthal directions. The initial conditions consisted of the numerically computed laminar CAF, which was disturbed by exciting all Fourier modes with small random noise.}  After initial transients, the amplitude of the imposed disturbances either increased or decreased exponentially in time and growth rates could be extracted { by using  exponential fits to the time series of mode amplitudes. This procedure is equivalent to using the power method to compute the leading eigenvalue}. The time step size $\Delta t$ was held  small (between $2\times10^{-6}$ and $2\times10^{-5}$ depending on $Pe$ and $Cn$) to ensure that temporal discretization errors were negligible. { The radial resolution was kept large enough (with $7$-$8$ points in the interface region)  to ensure that spatial discretization errors were also negligible.} This enabled a study of the convergence to the sharp-interface limit as a function of $Cn$ and $Pe$ only.

\begin{figure}[!ht]
        \centering
        \begin{subfigure}[b]{0.98\textwidth}
            \centering
		     \caption%
            {{\small}} 
            \includegraphics[width=\textwidth]{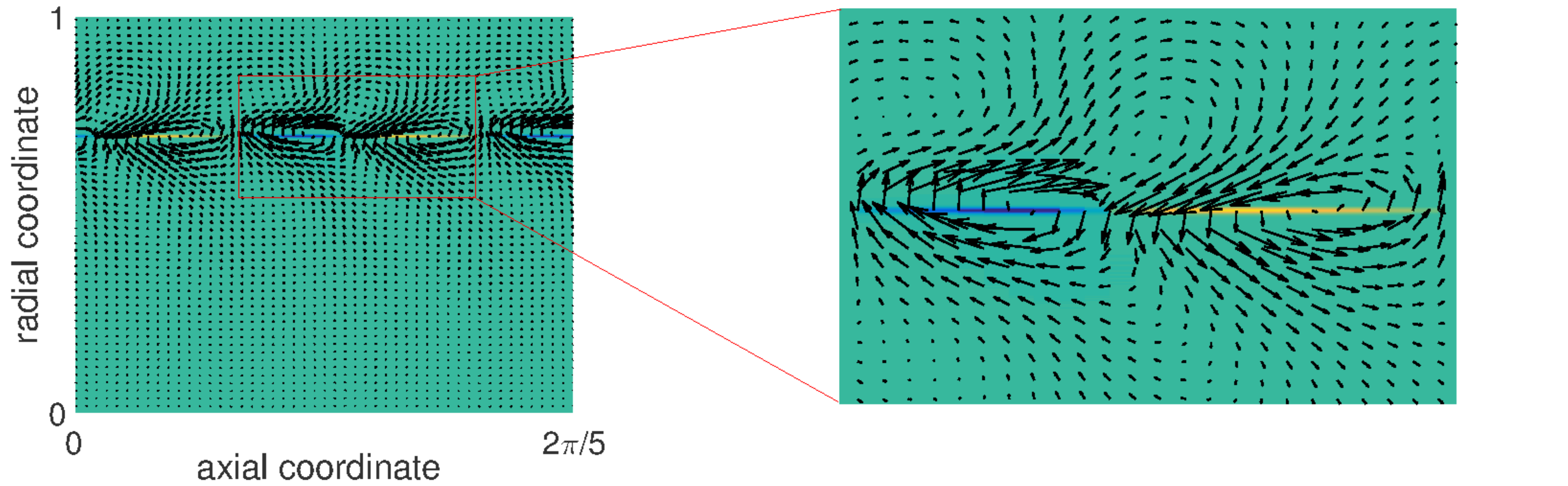}
   
            \label{fig:CAF1_disturbance}
        \end{subfigure}
        \hfill
        \begin{subfigure}[b]{0.98\textwidth}  
            \centering 
			\caption[]%
		    {{\small}}   
            \includegraphics[width=\textwidth]{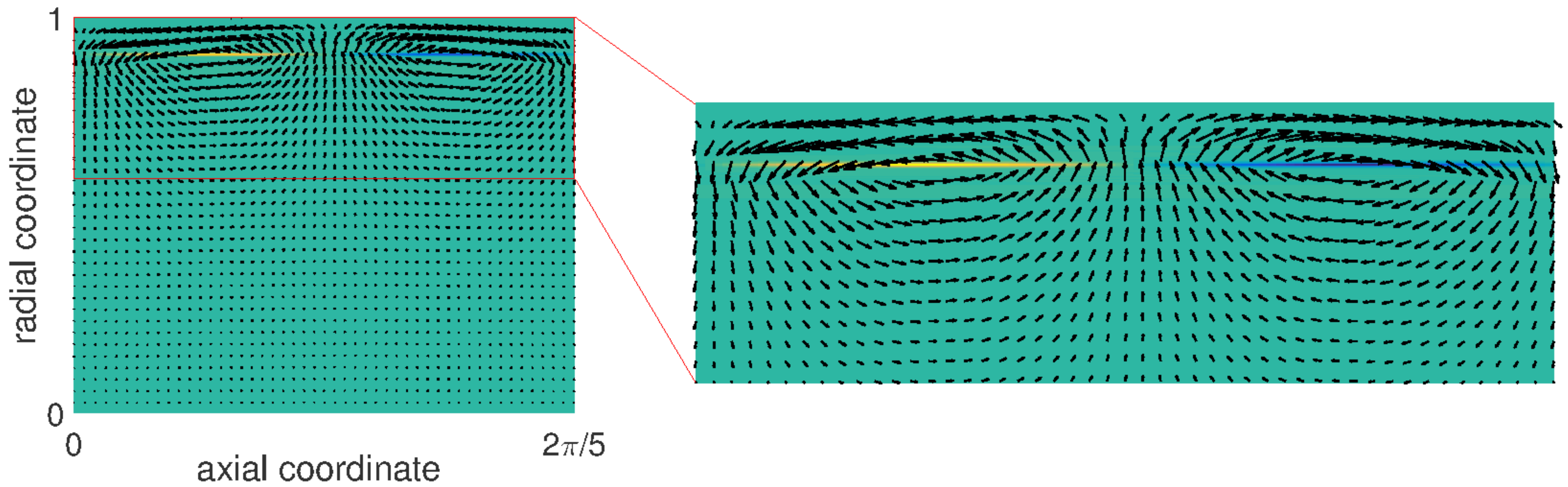}
            \label{fig:CAF2_disturbance}
        \end{subfigure}
\caption[holdup_ratio]
{\label{fig:optimal_modes} Velocity field (vectors) and phase field $C$ (colormap) of the most unstable eigenmode for (a) CAF1 and (b) CAF2 in table~\ref{tab:test_cases}. { Here $k_0=5$, corresponding to a pipe length of $L_z=2\pi/5\approx1.26$. For this pipe length, the dominant mode has axial wavenumber $5$ ($k=1$) in CAF1 and $10$ ($k=2$) in CAF2, respectively.}}
    \end{figure}

\subsection{CAF1: Instability driven by interfacial-friction}

\begin{figure}
\centering	
\includegraphics[width=0.60\linewidth]{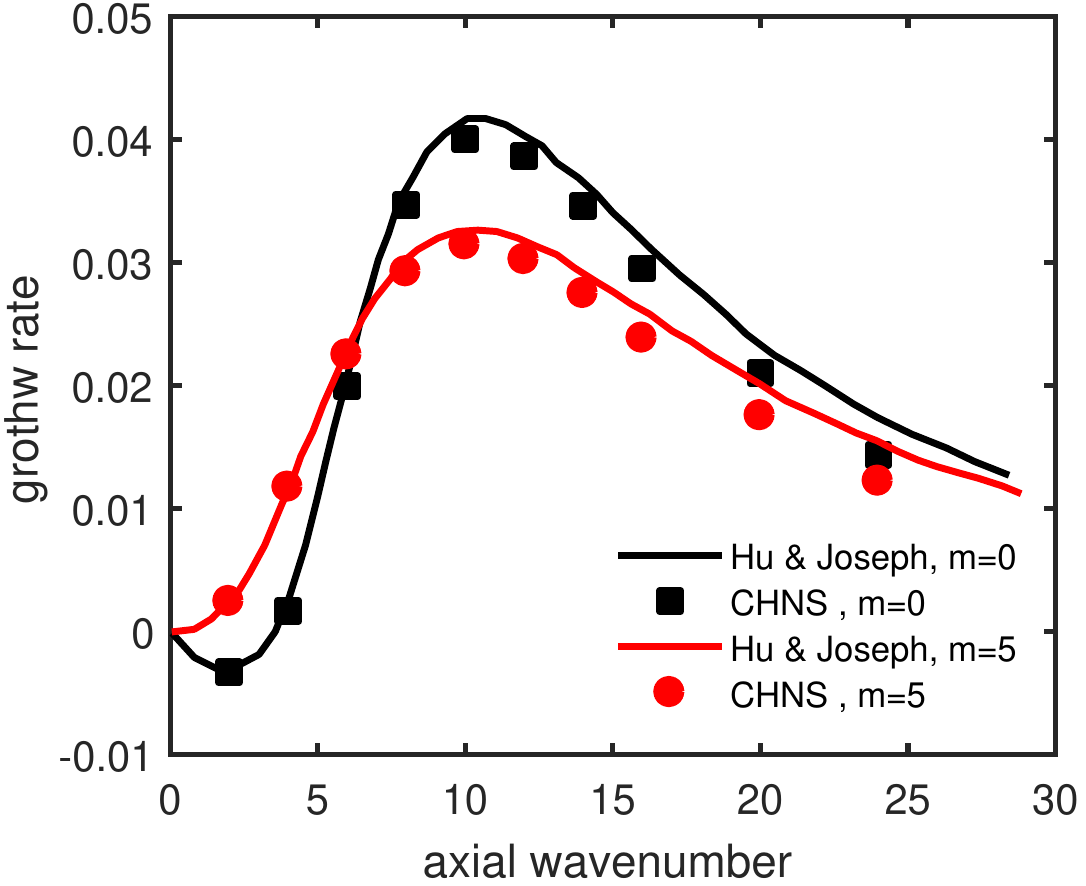}
\caption[growth rate comparison]{\label{fig:linstab_k}
Growth rate of axisymmetric ($m = 0$, black) and non-axisymmetric ($m = 5$, red) perturbations for CAF1. Our results (symbols) are compared with those reported in figure~1 of Hu and Joseph~\cite{Hu1989} (solid lines) in the sharp-interface limit. (For interpretation of the references to colour in this figure legend, the reader is referred to the web version of this article.)} 
\end{figure}

In this test case, there is no surface tension and $C$ enters the Navier--Stokes equations only via the variable viscosity. {Following Hu \& Joseph~\cite{Hu1989}, two cases were considered to validate our code: axisymmetric ($m=0$) and non-axisymmetric ($m=5$) perturbations to the base flow. }  Figure~\ref{fig:optimal_modes}a shows the fluid velocity fields (arrows) and phase fields $C$ (colormap) of the most unstable perturbation { ($m = 0$, according to~\cite{Hu1989})} with axial wavenumber $k=10$. As there is no capillary force and the Reynolds number is small, $Re=37.7$, the instability is driven here by interfacial friction. { A similar instability was also investigated in recent CHNS simulations of layered two-phase channel flow without surface tension \cite{Sahu2016}}. Figure~\ref{fig:linstab_k} shows the growth rate of the most dangerous disturbance, computed with $Cn=0.001$, $Pe=10^3$ and $N=288$ radial points, as a function of its axial wavenumber.  Our results are in very good agreement with the sharp-interface computations of Hu \& Joseph~\cite{Hu1989} both for axisymmetric ($m=0$) and non-axisymmetric ($m=5$) disturbances. As the axial wavenumber $k$ increases, the difference between our simulation and \cite{Hu1989} grows. Larger values of $k$ imply finer structures and therefore more radial finite difference points would be necessary to keep the same level of accuracy.

\begin{figure}[!ht]
        \centering
        \begin{subfigure}[b]{0.475\textwidth}
            \centering
             \caption%
            {{\small}}   
            \includegraphics[width=\textwidth]{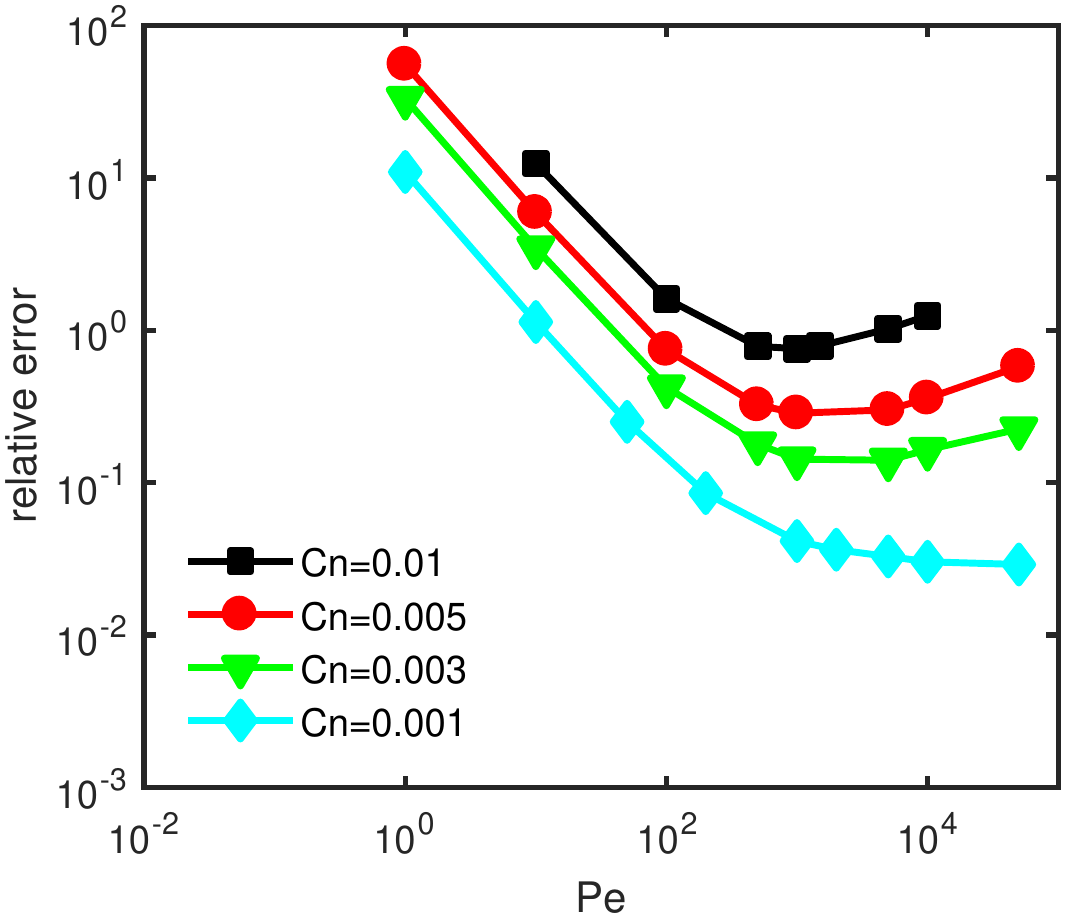}
            \label{fig:conv_Pe_CAF1}
        \end{subfigure}
        \hfill
        \begin{subfigure}[b]{0.475\textwidth}  
            \centering 
            \caption[]%
            {{\small}}   
            \includegraphics[width=\textwidth]{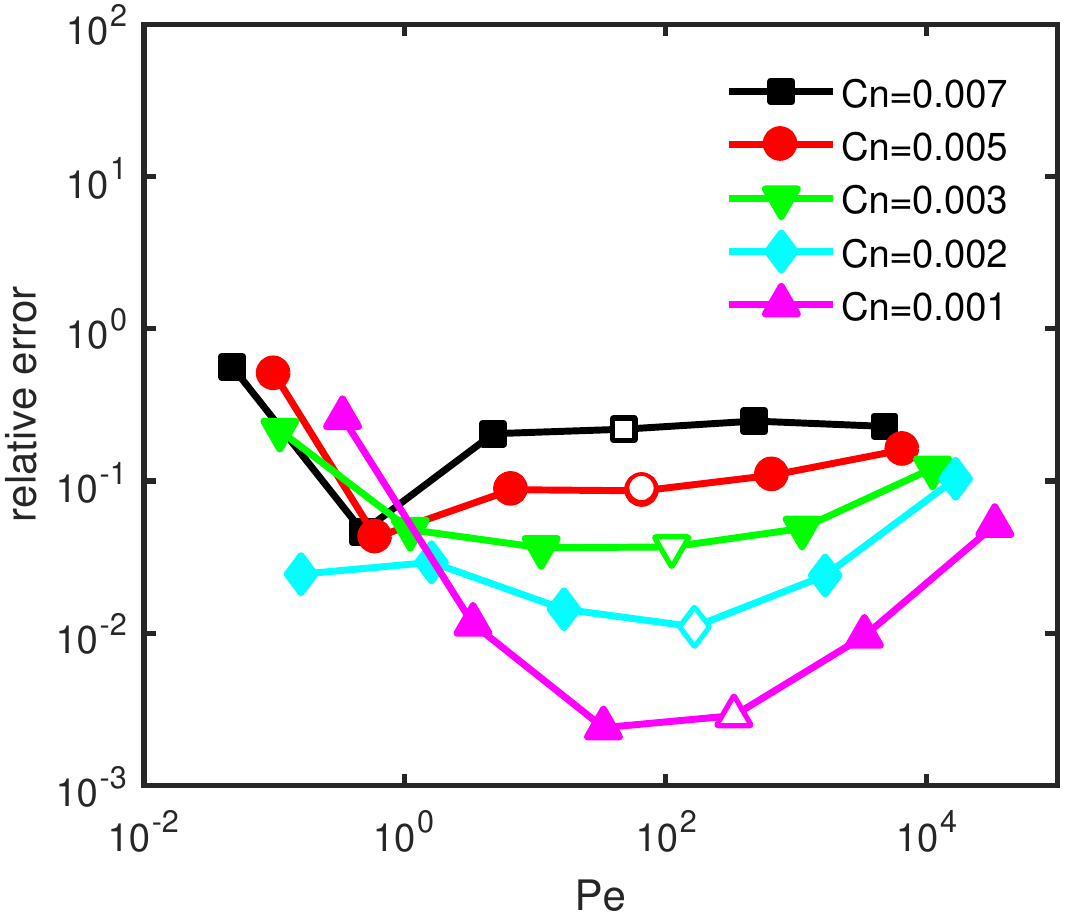} 
            \label{fig:conv_Pe_CAF2}
        \end{subfigure}
        \caption[base flow convergence]
        {\label{fig:growthrate_Pe} Relative error of the growth rate in our simulations with respect to that of the sharp interface solution (see table~1 in \cite{Hu1989}) as a function of $Pe$ for (a)  CAF1 and (b) CAF2. The axial wavenumber of the disturbance is $k=10$ in (a) and $k=5$ in (b). The open symbols in (b) highlight the relative errors obtained with $Pe_\text{opt}=1/(3\,Cn)$.}
    \end{figure}

\begin{figure}[!ht]
        \centering
        \begin{subfigure}[b]{0.475\textwidth}
            \centering
			\caption%
            {{\small}}  
            \includegraphics[width=\textwidth]{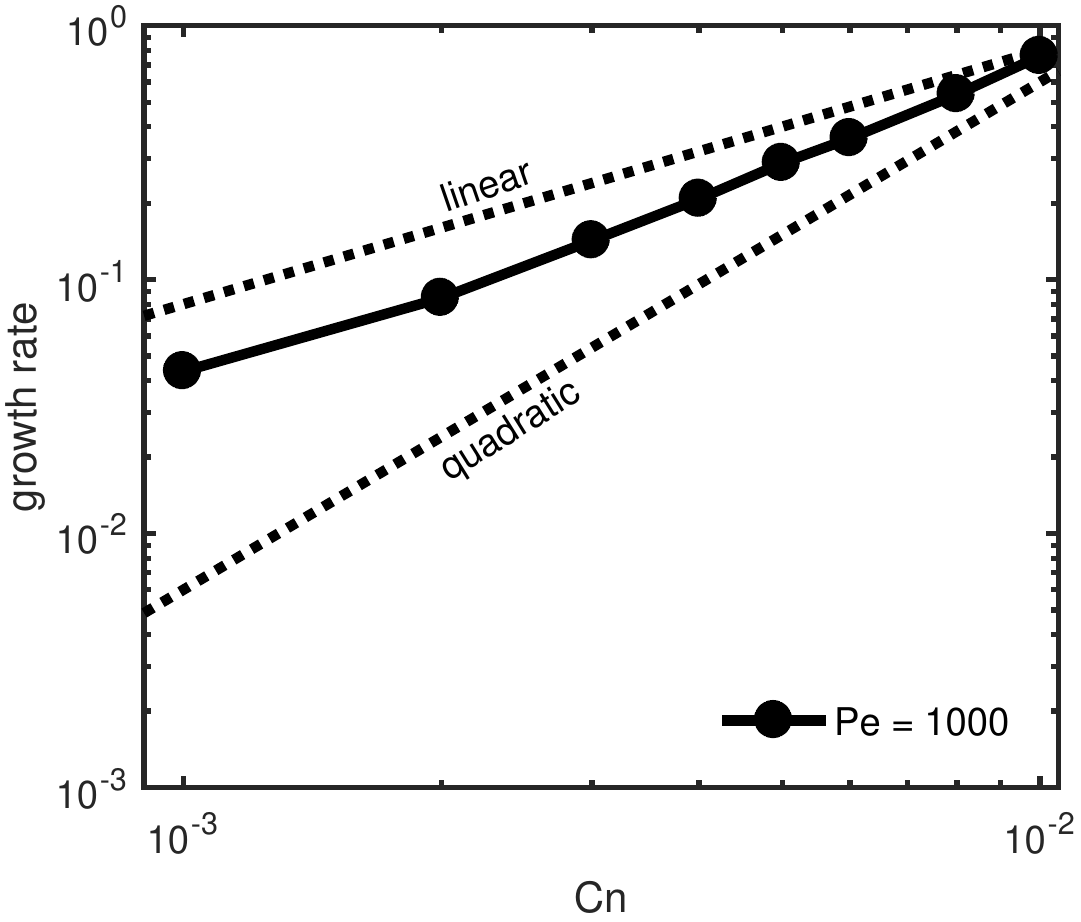}  
            \label{fig:convergence_Cn_CAF1}
        \end{subfigure}
        \hfill
        \begin{subfigure}[b]{0.475\textwidth}  
            \centering  
            \caption[]%
            {{\small}}  
            \includegraphics[width=\textwidth]{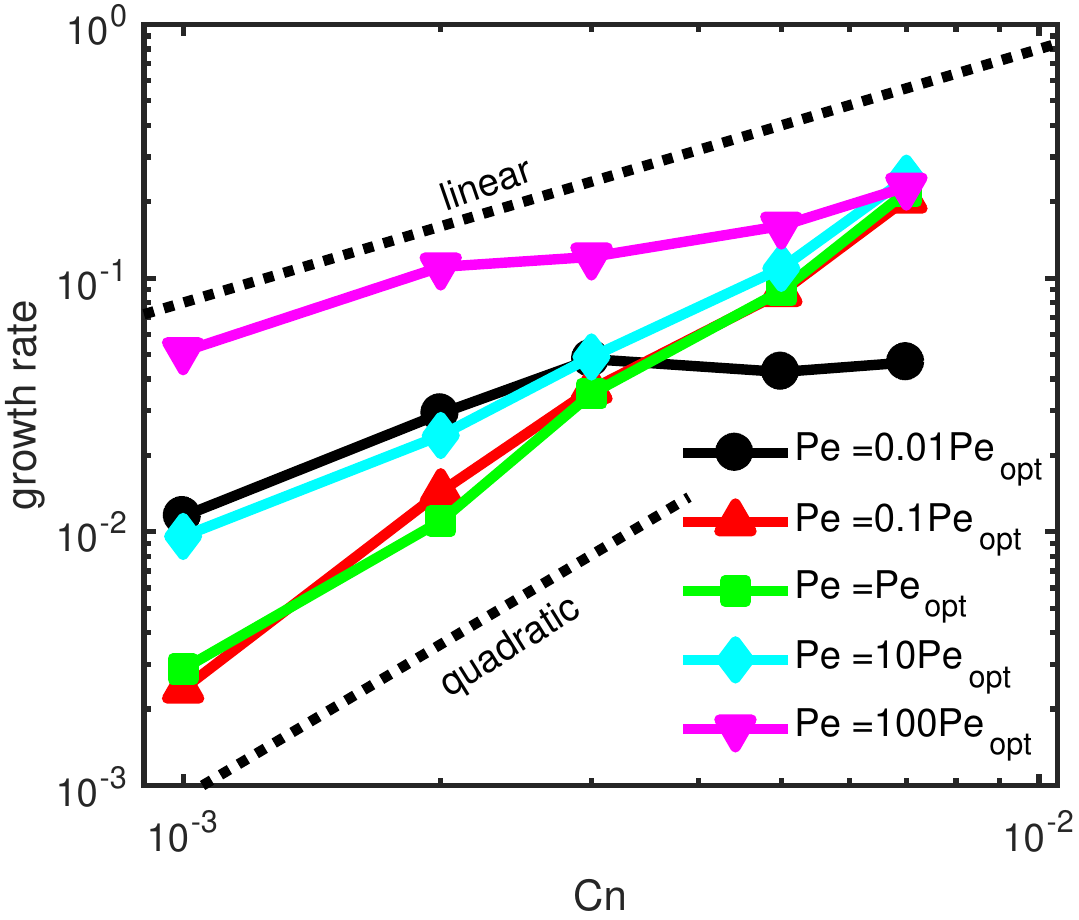}  
            \label{fig:convergence_Cn_CAF2}
        \end{subfigure}
\caption[growth rate comparison]{\label{fig:growthrate_convergence_Cn}
Convergence of the growth rate for  (a) CAF1 and (b) CAF2 from our simulations to the sharp interface solution~\cite{Hu1989} as $Cn$ is decreased. Linear and quadratic convergences are depicted with dotted thin lines. }%
    \end{figure}
    
According to the theory of Magaletti \emph{et al}.\ \cite{Magaletti2014}, in the absence of surface tension there is no optimal Peclet number and the correct interfacial dynamics should be recovered for sufficiently large $Pe$. Figure~\ref{fig:growthrate_Pe}a shows the relative error between the growth rate obtained in our simulations and that reported by Hu and Joseph~\cite{Hu1989} for $k=10$. In all cases, the relative error decreases initially as $Pe^{-1}$ when $Pe$ is increased from $O(1)$ to $O(10^3)$ and then reaches a plateau. If $Pe$ is further increased, the relative error increases monotonically, which differs from the theoretical arguments of~\cite{Magaletti2014}. The exception to this is the smallest $Cn=0.001$ considered, where the relative error seems to converge towards a constant value as $Pe$ is increased. This suggests that in this case the prediction of~\cite{Magaletti2014} might be valid only very close to the sharp-interface limit. Overall, our results suggest that in practice $Pe=10^3$ is sufficient for reproducing the dynamics without surface tension over the range of $Cn$ considered. In fact, figure~\ref{fig:convergence_Cn_CAF1} shows that by fixing $Pe=10^3$ a faster than linear convergence rate  is obtained, as $Cn$ is decreased. 

\subsection{CAF2: Instability driven by inertia}

In this test case, $Re=500$, $\hat\mu=1/20$ and the instability is predominantly inertial. Figure \ref{fig:CAF2_disturbance} shows that the velocity field of the disturbance is most intense in the annular region, in agreement with Hu and Joseph~\cite{Hu1989}, who showed that production is negligible with respect to dissipation in the core fluid because of the large viscosity. Figure~\ref{fig:growthrate_Pe}b shows that in the presence of surface tension there is an intermediate range of $Pe$ for which the relative error is minimized. { This minimum, which manifests itself clearly only as $Cn$ becomes sufficiently small ($Cn\lesssim 0.005$), is rather broad. It extends over one order of magnitude around the value of  $Pe_\text{opt}=1/(3Cn)$ proposed by Magaletti \emph{et al}.~\cite{Magaletti2014} as optimal. They proposed the pre-factor $1/3$ from their capillary-wave simulations with matched density and viscosity in a rectangular domain. Our results indicate that $1/3$ also works well in systems with cylindrical geometry, large viscosity ratio, strong inertia and shear}. When $Pe$ is decreased linearly, with pre-factors chosen in the near-optimal range, the convergence to the sharp-interface limit is for CAF2 quadratic in $Cn$ (see figure~\ref{fig:convergence_Cn_CAF2}).  The nearly quadratic convergences agrees with Jacqmin \cite{Jacqmin1999}, who pointed out that in practice the phase-field method may be faster than linear as predicted theoretically. 

\section{Water-lubricated {pipe flow}}
\label{sec:bamboo_wave}

\begin{table}
\centering
\begin{tabular}{|c|c|c|c|c|c|c|c|c|c|c|}
\hline
 Test case & $Re$ & $Ca$ &  $\hat\mu$ & $\hat\rho$ &
              $\hat V$ & $Fr$ & $f$ &$L$\\ 
\hline
CAF3 & 1.19 & 11.7  & 1/601 & 1.1  &  1.566 & 0.59 & 1.7455 & 2.45\\
\hline
\end{tabular}
\caption{\label{tab:CAF3} Dimensionless parameters of CAF in the nonlinear regime (bamboo wave) studied experimentally by Bai \emph{et al}.~\cite{Bai1992} and numerically by Li and Renardy~\cite{Li1999}. The pipe radius was $R=0.476$ cm, whereas $\rho_{\text{oil}}$=0.905 g/cm$^3$, $\rho_{\text{water}}$=0.995 g/cm$^3$, $\mu_{\text{oil}}=6.01$ dyn$\cdot$s/cm$^2$, $\mu_{\text{water}}=0.01$ dyn$\cdot$s/cm$^2$ and $\sigma=8.54$ dyn/cm. From these experimental parameters, Li \& Renardy \cite{Li1999} inferred the position of the interface  of the core oil flow ($\eta=0.78125$) and the driving pressure gradient ($f=1.7455$) for their simulations. 
} 
\end{table}

Bai \emph{et al}.~\cite{Bai1992} investigated experimentally CAF of very viscous oils lubricated with water ($\hat\mu=1/601$, $\hat\rho=1.1$) in vertical pipes and revealed a wide variety of flow patterns depending on the flow rates of the two phases. In this section, we focus on a case in which the flow is driven upward against gravity with a constant pressure gradient and closely follow the procedure employed by Li and Renardy~\cite{Li1999} in their VoF-simulations of this case (see table~\ref{tab:CAF3} for the flow parameters). We first obtained the laminar CAF at the corresponding parameters, then determined its linear instability and finally computed the resulting nonlinear flow pattern.

\subsection{Linear stability of laminar core-annular flow}

\begin{figure}[!ht]
\centering	
\includegraphics[width=0.6\textwidth]{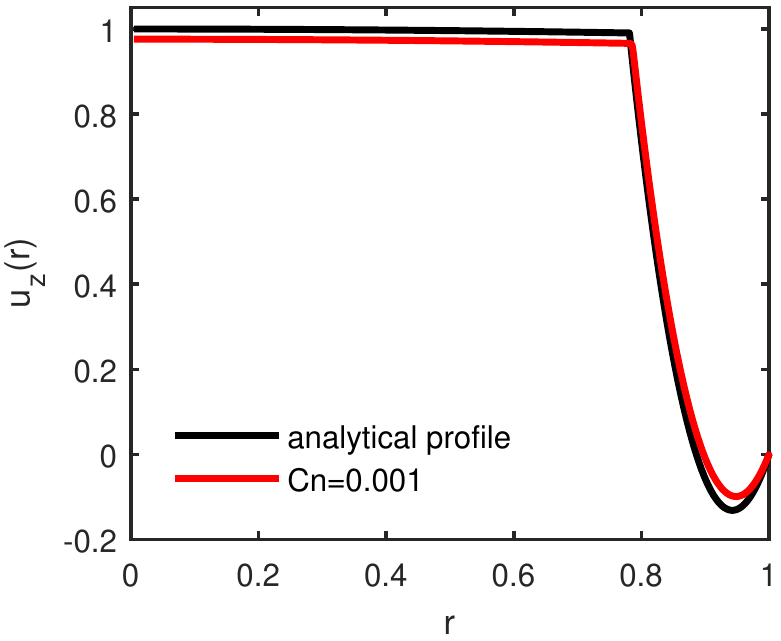}
\caption[basic flow up-flow]
{\label{fig:basicflow_up_flow} Laminar CAF in the sharp-interface limit (black line) and for $Cn=0.001$ (red line). In both cases the total volume flux is equal, whereas the driving pressure gradient is $f=1.7455$ and  $1.7479$, respectively. The flow parameters are given in table~\ref{tab:CAF3}. (For interpretation of the references to colour in this figure legend, the reader is referred to the web version of this article.)}
\end{figure}

Li and Renardy~\cite{Li1999} estimated from the experiments of Bai \emph{et al}.~\cite{Bai1992} that the dimensionless pressure gradient was $f=1.7455$. In this paper, the value of the pressure gradient was chosen so that the same volume flux as for the laminar CAF in the sharp-interface limit was obtained. This yielded $f=1.7479$ for $Cn=0.001$, obtained with $N=576$ radial points. Figure~\ref{fig:basicflow_up_flow} shows that the computed laminar CAF renders a reasonable approximation of the sharp-interface solution, with 7.5\% maximum error with respect to eq.~\eqref{equ:basic_flow}, despite the very high viscosity ratio of the fluids used in the experiments. 

\begin{figure}
        \centering
        \begin{subfigure}[b]{0.475\textwidth}
            \centering
             \caption%
            {{\small}}  
            \includegraphics[scale=0.6]{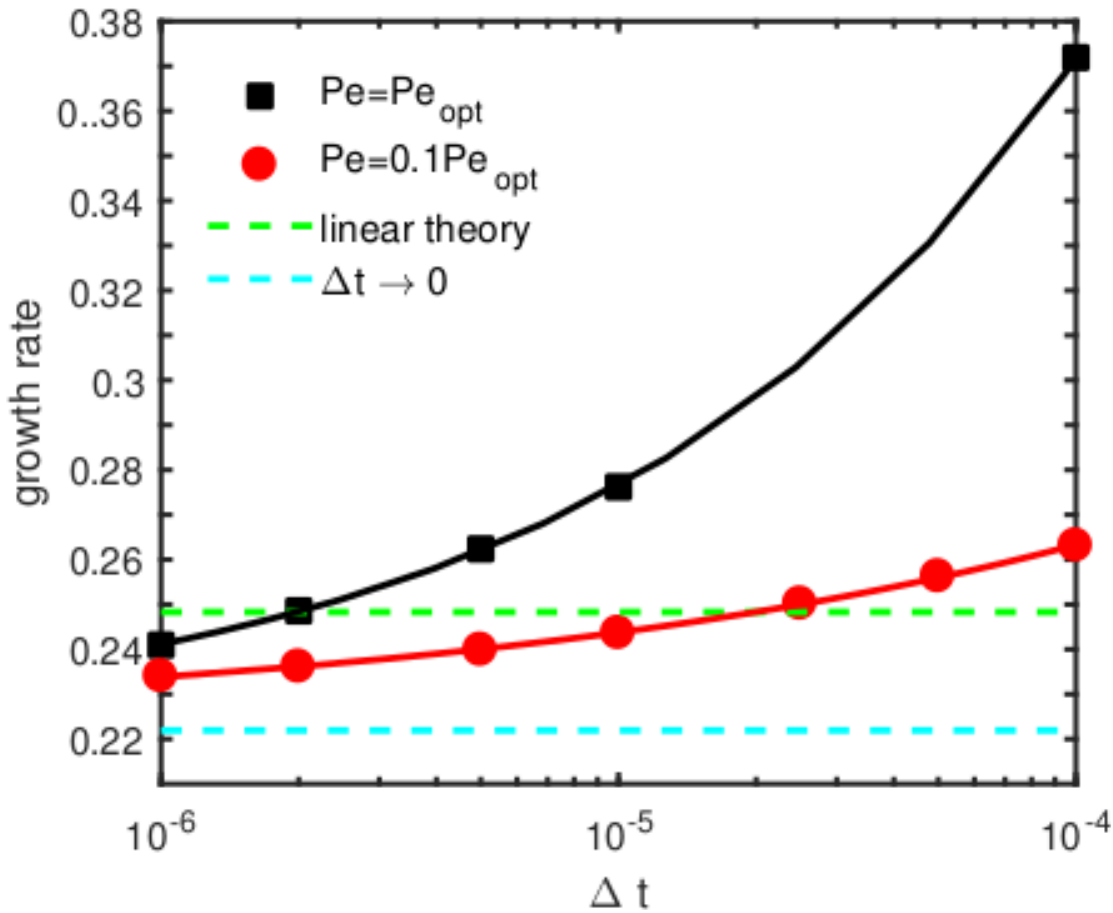}            
            \label{fig:growth_rate_CAF3}
        \end{subfigure}
        \hfill
        \begin{subfigure}[b]{0.475\textwidth}  
            \centering 
             \caption[]%
            {{\small}}    
            \includegraphics[scale=0.6]{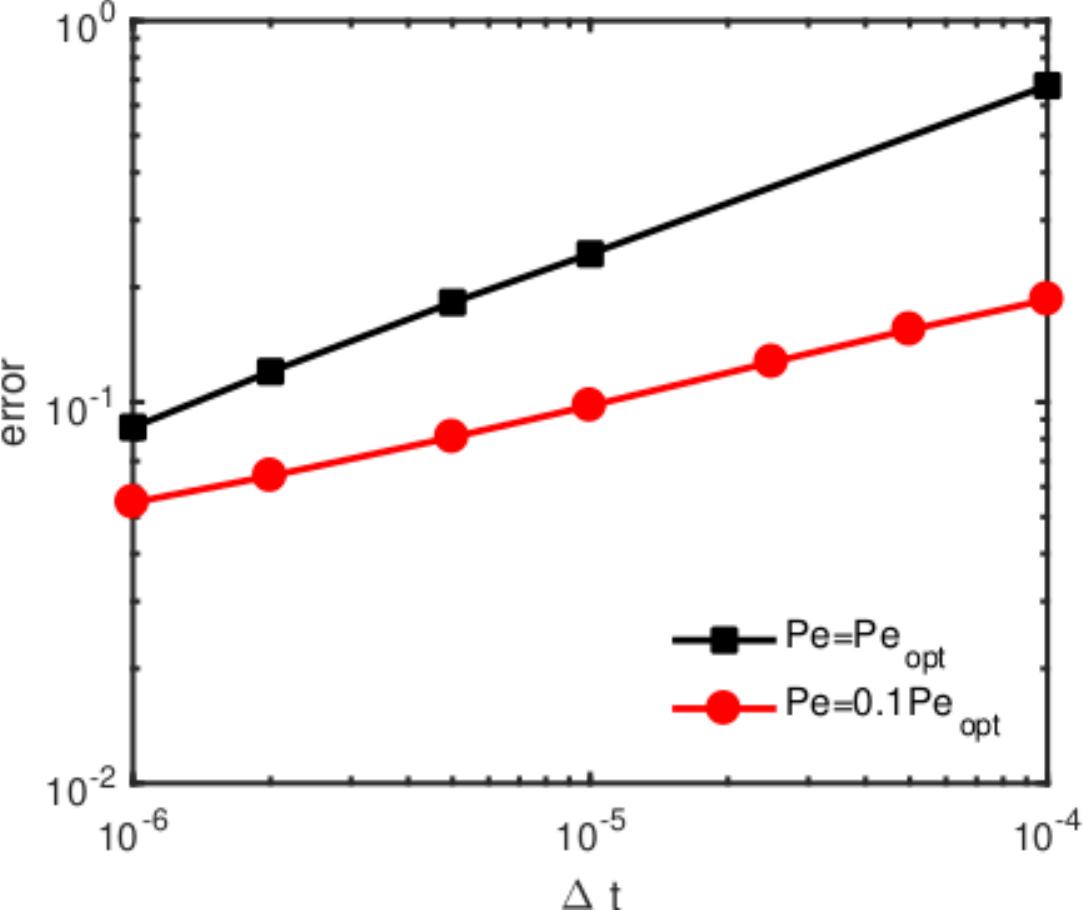}
            \label{fig:error_CAF3}
        \end{subfigure}
\caption[bamboo wave Pe_dt]
{\label{fig:Pe_dt_upflow} (a) Growth rate of the linear instability of the laminar CAF (shown as a red line in figure~\ref{fig:basicflow_up_flow}) as a function of $\Delta t$ with $Pe=Pe_\text{opt}$ (\mysquare{black}) and $Pe=0.1Pe_\text{opt}$ (\mycircle{red}). Power law fits of the data of the form $a\cdot \Delta t^b+c$, where $c=0.22$ (cyan dashed line) approximates the growth rate in the limit $\Delta t \rightarrow 0$, are shown as solid lines. The fit parameters are $a=8.02$, $b=0.43$ for $Pe=Pe_\text{opt}$ and $a=0.59$, $b=0.29$ for $Pe = 0.1Pe_{opt}$. The green dashed line (0.248) depicts the growth rate of the sharp-interface limit~\cite{Li1999}. (b) Convergence with respect to $\Delta t$ for $Pe=Pe_\text{opt}$ and $0.1Pe_\text{opt}$, taking $c=0.22$ as reference. (For interpretation of the references to colour in this figure legend, the reader is referred to the web version of this article.)
}    
\end{figure}

Following Li and Renardy~\cite{Li1999}, we destabilized the laminar CAF by disturbing only the first axisymmetric mode in a pipe of length $L=\pi/1.28\approx 2.45$. The computed growth rate as a function of the time-step size is shown in Figure~\ref{fig:Pe_dt_upflow}a for $Cn=0.001$. The behavior of the temporal error was examined by performing simulations with two different Peclet numbers, $Pe=Pe_\text{opt}=333$ and $Pe=0.1Pe_\text{opt}=33.3$. Both cases converged towards the same value as $\Delta t\rightarrow0$ { (within a discrepancy of 0.7\%), which was calculated by fitting a power law to the finite $\Delta t$ results. However,  for $0.1Pe_\text{opt}$  and $\Delta t=10^{-4}$, smaller error than for $Pe=Pe_\text{opt}$ and $\Delta t =10^{-5}$ was achieved (see figure~\ref{fig:Pe_dt_upflow}b).} Clearly, $0.1Pe_\text{opt}$ allows a much larger time-step size for the same accuracy, so we fixed $Pe=0.1Pe_\text{opt}$ in the following nonlinear analysis. We note that our result differs by 10\% with respect to the sharp-interface limit~\cite{Li1999}. Hence it can be concluded that the error in the linear stability analysis stems mainly from the diffuse-interface approximation of the underlying laminar CAF. 

The leading eigenvalue of the instability has non-zero imaginary part, and corresponds to a Hopf bifurcation of the laminar CAF. Because of the mean positive advection in the vertical direction, the bifurcating flow pattern is generically expected to be an upward-traveling wave in the axial direction, satisfying
\begin{equation}\label{eq:tw}
 \boldsymbol u (r,z,\theta,t) =  \boldsymbol u (r,z- c\,t,\theta,0),
\end{equation}
with $c>0$. The computed linear wave speed $c=0.83$ was found in good agreement with the linear stability analysis of the sharp-interface limit \cite{Li1999}, with $c=0.85$.  

\subsection{Nonlinear bamboo waves}

\begin{figure}
        \centering
        \begin{subfigure}[b]{0.9\textwidth}
            \centering
            \caption%
            {{\small }}    
            \includegraphics[width=\textwidth]{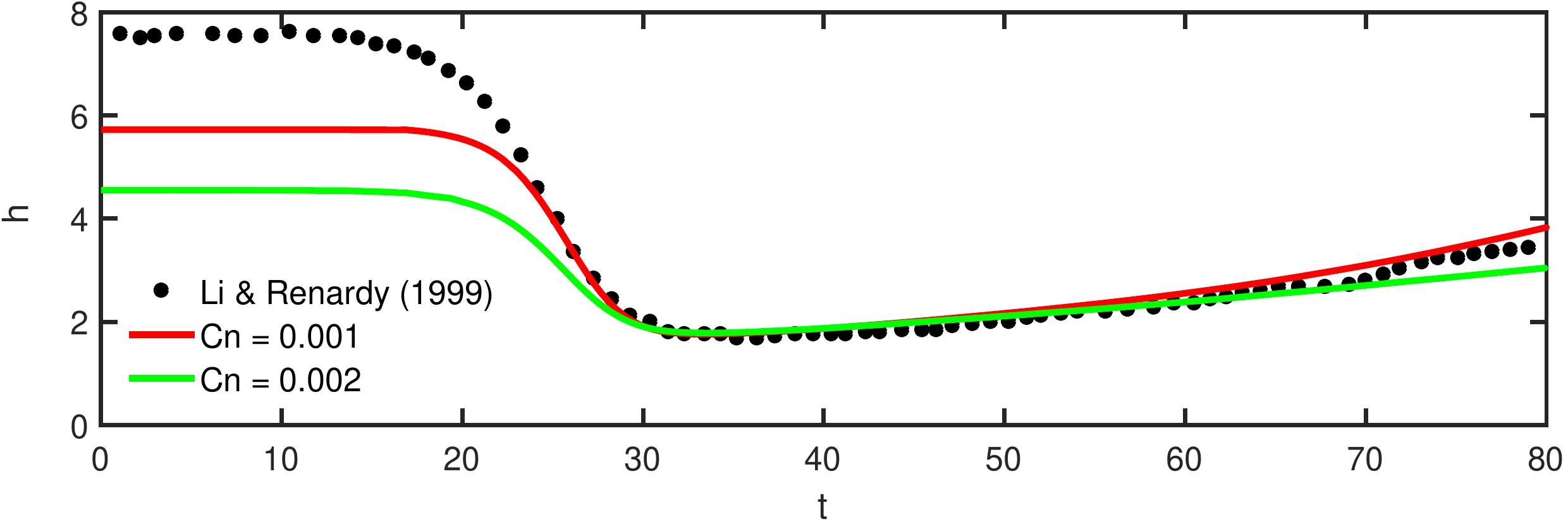}
            \label{fig:CAF3_velocity}
        \end{subfigure}
        \hfill
        \begin{subfigure}[b]{0.90\textwidth}  
            \centering 
            \caption[]%
            {{\small}}    
            \includegraphics[width=\textwidth]{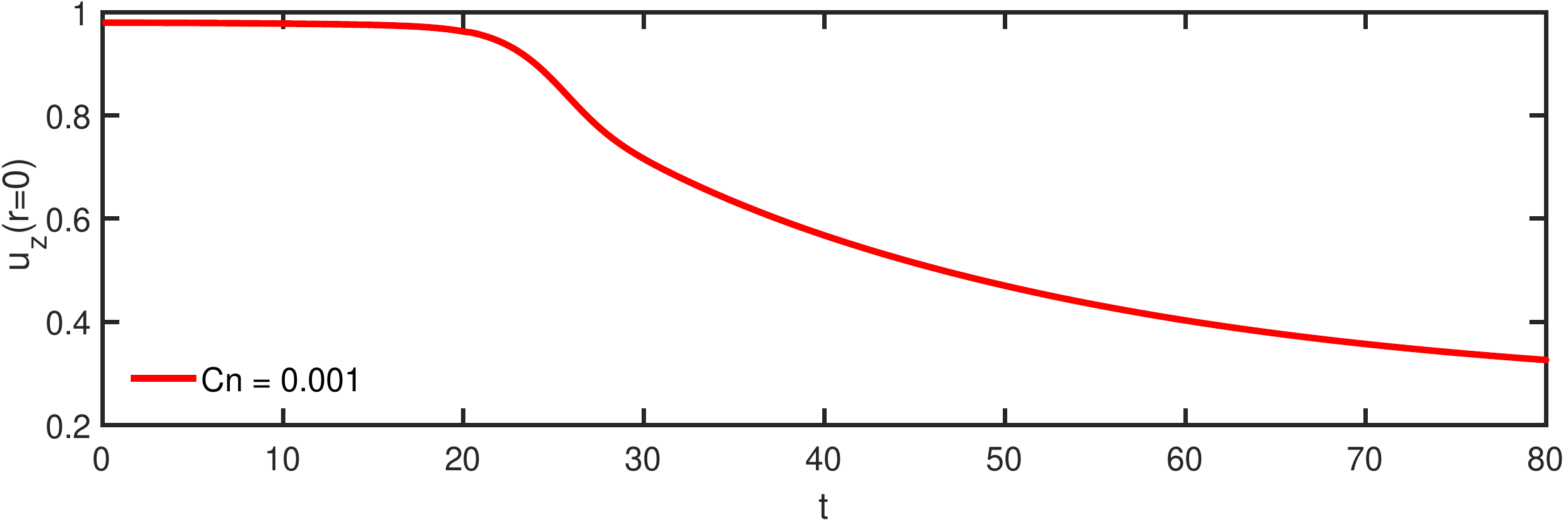}
             \label{fig:CAF3_holdup}
        \end{subfigure}
\caption[holdup_ratio]
{\label{fig:holdup_ratio} Onset of bamboo waves with $Cn=0.001$  and the flow parameters given in table~\ref{tab:CAF3}. Temporal evolution of (a) the hold-up ratio, eq.~\eqref{eq:holdup} and (b) the centerline velocity of the stream-wise averaged flow profile. The hold-up ratio computed with $Cn=0.002$ is shown as a green line, whereas the computations of Li and Renardy~\cite{Li1999} are shown as black circles. Their data were shifted in time so that the nonlinear evolution of both simulations matches. This was necessary because it was not possible to reproduce their perturbation, which led to slightly different transition dynamics. (For interpretation of the references to colour in this figure legend, the reader is referred to the web version of this article.)}
    \end{figure}

\begin{figure}[!ht]
\centering	
		\begin{subfigure}[b]{0.18\textwidth}
            \centering
             \caption%
            {{\small }} 
            \begin{tikzpicture}
            \node[anchor=south west,inner sep=0] (image) at (0,0) 			{\includegraphics[width=\textwidth, height=5.8cm]{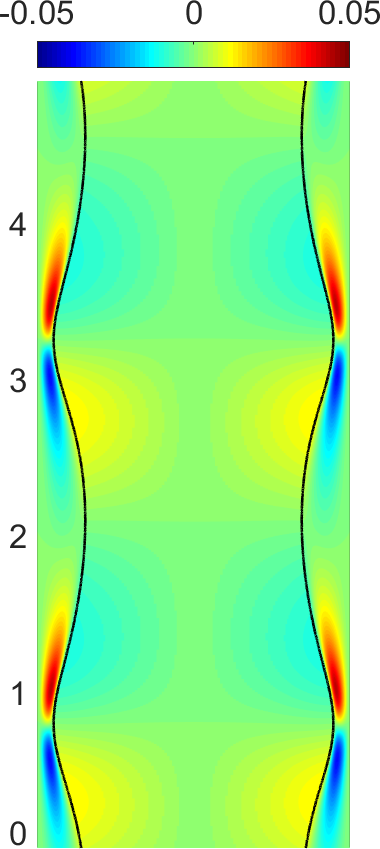}};
            \end{tikzpicture}%
            \label{fig:bamboo_wave_vr}
        \end{subfigure}
        \begin{subfigure}[b]{0.18\textwidth}
            \centering
			 \caption[]%
            {{\small }}               
            \begin{tikzpicture}
            \node[anchor=south west,inner sep=0] (image) at (0,0) 			{\includegraphics[width=\textwidth, height=5.8cm]{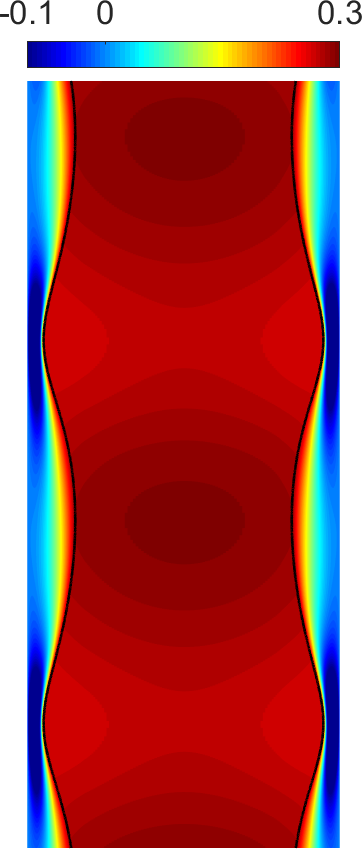}};
            \end{tikzpicture}%
            \label{fig:bamboo_wave_vz}
        \end{subfigure}
        \begin{subfigure}[b]{0.18\textwidth}  
            \centering
          \caption[]%
            {{\small }}    
            \begin{tikzpicture}
            \node[anchor=south west,inner sep=0] (image) at (0,0) 			{\includegraphics[width=\textwidth, height=5.8cm]{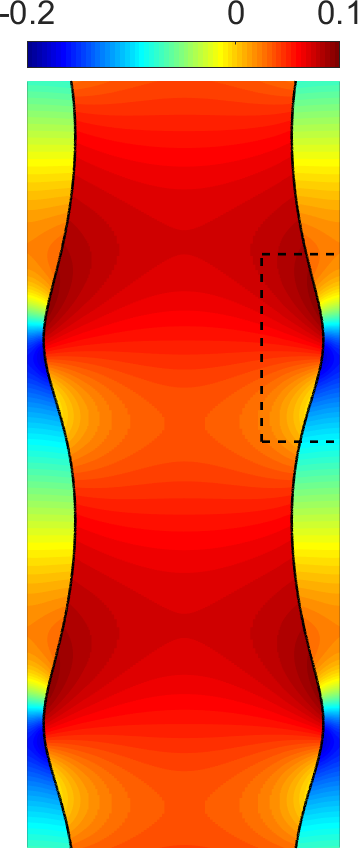}};
            \end{tikzpicture}%
              \label{fig:bamboo_wave_p}
        \end{subfigure}
       \begin{subfigure}[b]{0.18\textwidth}  
 			\caption[]%
            {{\small}}  
		\begin{tikzpicture}
		\node [anchor=west] (start) at (2.8,1.5) {\Large  };
		\node [anchor=west] (g) at (3.1,.40) {\Large $\vec{g}$ };
		\begin{scope}
	    \node[anchor=south west,inner sep=0] (image) at (0,0) 			{\includegraphics[width=\textwidth, height=5.74cm]{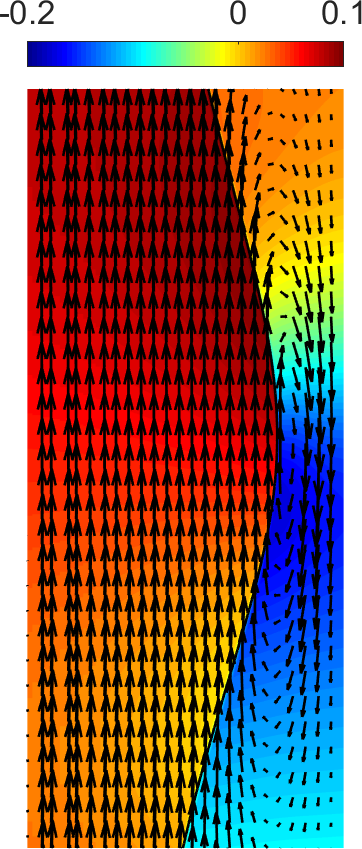}};
		\draw [-stealth, line width=3pt, black] (start) -- ++(0.0,-1.5);
		\end{scope}
\end{tikzpicture}%
            \label{fig:bamboo_wave_pvec} 
        \end{subfigure}
\caption[bamboo_wave]
{\label{fig:bamboo_wave} Nonlinear bamboo waves at $t=84$ computed with $Cn=0.001$ for the flow parameters in  table~\ref{tab:CAF3}. Two wavelengths of the computed pattern are shown. The bold black line shows the position of the interface between oil and water, which is represented as the contour $C=0$. (a) Colormap of the radial (wall-normal) velocity component $u_r$ in the $r$-$z$ cross-section. (b) Colormap of streamwise velocity $u_z$. (c) Colormap of the fluid pressure $p$. The linear pressure profile driving the flow, $p_f=-f\,z$, is not shown. (d) Velocity field $(u_r,u_z)$ and pressure (colormap) near the wave crest in the small region enclosed by the window defined by dashed lines in panel (c).}
\end{figure}

Figure~\ref{fig:holdup_ratio}a illustrates the temporal evolution of the flow instability starting from a slightly disturbed laminar CAF. At around $t=20$ nonlinear effects begin to kick in and the characteristic bamboo-wave pattern shown in figure~\ref{fig:bamboo_wave} fully emerges by $t=30$. { Clearly, the interface between the two fluids ceases to be parallel to the pipe axis, and a large number of axial modes is required to accurately resolve the profile of $C$ across the interface.} Here, $2K=3456$ axial Fourier modes were used (the number of radial points was not changed, $N=576$). Because of the high computational cost of { these two-dimensional} nonlinear simulations, we set $\Delta t=2\times10^{-5}$  and $Pe = 33.33$, which gave a good compromise between computing cost and temporal error (see figure~\ref{fig:Pe_dt_upflow}). 

It is worth noting that the agreement between { the hold-up ratio} in our simulations and those of Li and Renardy~\cite{Li1999} is excellent during the transition and the nonlinear saturation phase, whereas the discrepancy in the laminar phase is much larger (see figure~\ref{fig:holdup_ratio}a). { In the laminar flow, the interface and the fluid velocity are parallel to each other. In the Cahn--Hilliard approximation, the interface is smooth and because of the high viscosity ratio, the magnitude of the velocities in the oil and the water phases is lower than in the analytic sharp-interface solution (see figure~\ref{fig:basicflow_up_flow}). This results in a significantly lower hold-up ratio. Although eq.~\eqref{eq:err_lam} shows that this could be ameliorated by further decreasing $Cn$, the computational cost would be enormous. By contrast, the VoF simulation of  Li \& Renardy~\cite{Li1999} can deal with a sharp interface and therefore does not suffer from this problem, rendering better agreement with the analytic solution. In the nonlinear regime, however, the interface is no longer parallel to the fluid velocity and the hold-up ratio of the bamboo pattern becomes less sensitive to the interface thickness.}  This was confirmed by doubling the Cahn number in our simulations (see  the green line in figure~\ref{fig:holdup_ratio}a), which worsened the agreement in the laminar flow phase, but did not alter the transition process significantly. 

The close-up of figure~\ref{fig:bamboo_wave}d shows that the oil (core) flows upward (from left to right), and part of the water (annular fluid) is trapped in the trough of the wave and is carried upward as well{, thereby decreasing the hold-up ratio}. The water near the wave crests exhibits swirls and is carried downward. The large gradients in the very thin gap between interface and pipe wall at the wave crest enhance the flow resistance. As a result the transport of oil is strongly reduced, as illustrated by a drop of the centerline velocity by nearly a factor of $3$ with respect to the laminar CAF (see figure~\ref{fig:holdup_ratio}b). 

Finally, we note that although the flow pattern changed litte after $t=30$, the flow continued to evolve until the end of the simulation. Clearly, neither the hold-up ratio nor the centerline velocity saturated at a constant value. Traveling waves satisfying eq.~\eqref{eq:tw} are relative equilibria, for which integral quantities, such as the hold-up ratio, must remain constant in time. This implies that the computed waves are not exact traveling waves, but have suffered further secondary bifurcations already. This is in qualitative agreement with Bai \emph{et al}.~\cite{Bai1992}, who noted that the bamboo waves observed experimentally were not completely axisymmetric and consisted of the superposition of waves of different wavelengths traveling at different speeds. Studying these phenomena in a realistic setup would require non-axisymmetric { (three-dimensional)} simulations with much longer domains than used here and is out of the scope of this work.

\section{Conclusions and outlook}

Phase-field methods have long been used to simulate multiphase flows, mostly to demonstrate their capability of dealing with topological changes, such as droplet break up and coalescence. By contrast, quantitative investigations of their accuracy have remained scarce to date, especially for realistic geometries and fluid properties. This was undertaken here with pseudo-spectral simulations of the Cahn-Hilliard-Navier-Stokes equations for two-phase pipe flow. We validated our code against solutions of the governing equations in the sharp-interface limit (i.e.\ the Navier--Stokes equations with stress and velocity boundary conditions at the interface between the two fluids). Our tests included an analytic laminar flow solution, linear stability analysis with axisymmetric and non-axisymmetric modes, and a fully nonlinear flow pattern. {In all cases, the relationship between interface thickness and mobility derived by  Magaletti \emph{et al}.\ \cite{Magaletti2014}, $Pe \propto Cn^{-1}$, was shown to  converge to the sharp-interface limit}. Our results suggest that a smaller pre-factor (e.g.\ $1/30$ instead of their proposed $1/3$) could allow the same accuracy with an order of magnitude larger time step. { In general, we expect the optimal pre-factor to be problem dependent.}

The setup chosen here exhibits moderate interface areas and deformations, and no topological changes. In this case, interface tracking methods are clearly superior. {For example, in laminar CAF the velocity is parallel to the interface and the model error of the CHNS increases as the viscosity ratio $\hat \mu$ decreases. In the limit of very low viscosity ratio, corresponding here to very viscous oils in the core, the error is $\varepsilon = \mathcal{O}(Cn \log \hat \mu^{-1})$, suggesting that phase-field methods are less suited than other methods, such as VoF, to simulate two-phase flows with very large difference in the viscosity of the two phases.} However, our choice of setup  was made in order to stringently test phase-field methods against well-known solutions and for flows that can be realized experimentally. Phase-field methods are specially well suited for the treatment of turbulent multiphase flows, where interfaces stretch, deform and even break, largely increasing its surface area \cite{Scarbolo2015}. For other interface-capturing methods, such as VoF or level set, which rely on the calculation of the local interface curvature, the additional computing cost cannot be  neglected. By contrast, the cost of the simulation with CHNS is independent of the shape and area of the interfaces, which makes CHNS especially attractive for large volume concentrations of the dispersed phase and for turbulent flows ~\cite{Roccon2019}. For the latter, one obvious requirement to faithfully capture the flow physics is that the interface thickness must be smaller than the Kolmogorov scale. Requirements on the accuracy with respect to time-step size remain however to be determined.

Finally, we note that CHNS enable the application of dynamical-system approaches to multiphase flows. In particular, CHNS allow the direct computation of nonlinear traveling waves satisfying eq.~\eqref{eq:tw}, and more complex solutions such as relative periodic orbits \cite{willis2013}, by solving \eqref{dimlessN-s}--\eqref{equ:CH_dimless} directly with the Newton method~\cite{kawahara2012}. {This seems to be more difficult} with other methods (such as VoF), because the underlying equations must be formulated as a system of smooth partial differential equations, such as \eqref{dimlessN-s}--\eqref{equ:CH_dimless}. Dynamical-system approaches have helped in elucidating the transition to turbulence in wall-bounded (single-phase) flows \cite{eckhardt2018}, and may prove useful to tackle the rich nonlinear dynamics exhibited by two-phase pipe flows \cite{Joseph1997}.

\section*{Acknowledgement}
We thank Dr. Ashley P. Willis for his help with our understanding of 
the influence matrix method and the singularity removal in polar coordinates.
We acknowledge the Deutsche Forschungsgemeinschaft (Project No. FOR 1182)
for financial support and acknowledge the Norddeutsche Verbund f\"ur Hoch- und H\"ochstleistungsrechnen (HLRN) for computing resources. 
\section*{References}
\bibliographystyle{plain}
\bibliography{Multiphase}

\begin{thebibliography}{10}

\bibitem{Ahmadi2016}
S.~Ahmadi, A.~Roccon, F.~Zonta, and A.~Soldati.
\newblock Turbulent drag reduction in channel flow with viscosity stratified.
\newblock {\em Computers and Fluids}, page 073302, 2015.

\bibitem{Anderson1998}
D.~M. Anderson, G.~B. McFadden, and A.~A. Wheeler.
\newblock Diffusive-interface methods in fluid mechanics.
\newblock {\em Ann.\ Rev.\ Fluid Mech.}, 30:139--65, 1998.

\bibitem{Bai1992}
R.~Bai, K.~Chen, and D.~D. Joseph.
\newblock Lubricated pipelining: stability of core-annular flow. part5.
  experiments and comparison with theory.
\newblock {\em J.\,Fluid Mech.}, 240:97--142, 1992.

\bibitem{Cahn1958}
J.~W. Cahn and J.~E. Hilliard.
\newblock Free energy of a nonuniform system. {I}. {I}nterfacial free energy.
\newblock {\em J. Chem. Phys.}, 28:258--267, 1958.

\bibitem{Cristini2004}
V.~Cristini and Y.~C. Tan.
\newblock Theory and numerical simulation of droplet dynamics in complex
  flows---a review.
\newblock {\em Lab Chip}, 4:257--264, 2004.

\bibitem{Dong2012}
S.~Dong and J.~Shen.
\newblock A time{-}stepping scheme involving constant coefficient matrices for
  phase-field simulations of two-phase incompressible flows with large density
  ratios.
\newblock {\em J.\,Comput.\ Phys.}, 213:5788–5804, 2012.

\bibitem{eckhardt2018}
Bruno Eckhardt.
\newblock Transition to turbulence in shear flows.
\newblock {\em Physica A: Statistical Mechanics and its Applications},
  504:121--129, 2018.

\bibitem{Govindarajan2014}
R.~Govindarajan and K.C. Sahu.
\newblock {Instabilities in viscosity-stratified flow}.
\newblock {\em Ann.\ Rev.\ Fluid Mech.}, 46:331--353, 2014.

\bibitem{Guseva2015}
A.~Guseva, A.~P. Willis, R.~Hollerbach, and M.~Avila.
\newblock {Transition to magnetorotational turbulence in Taylor--Couette flow
  with imposed azimuthal magnetic field}.
\newblock {\em New\ J.\ Phys.}, 17:093018, 2015.

\bibitem{Hirt1981}
C.~W. Hirt and B.~D. Nichols.
\newblock Volume of fluid (vof) method for the dynamics of free boundaries 1.
\newblock {\em J.\,Comput.\ Phys.}, 39:201–225, 1981.

\bibitem{Hu1989}
H.~H. Hu and D.~D. Joseph.
\newblock Lubricated pipelining : stability of core-annular flow. part 2.
\newblock {\em J.\,Fluid Mech.}, 205:359--396, 1989.

\bibitem{Jacqmin1999}
D.~Jacqmin.
\newblock {Calculation of two-phase Navier--Stokes flows using phase-field
  modeling}.
\newblock {\em J.\,Comput.\ Phys.}, 155:96--127, 1999.

\bibitem{Joseph1997}
D.~D. Joseph.
\newblock Core{-}annular flows.
\newblock {\em Annu.\ Rev.\ Fluid\ Mech.}, 29:65--90, 1997.

\bibitem{kawahara2012}
G.~Kawahara, M.~Uhlmann, and L.~van Veen.
\newblock {The significance of simple invariant solutions in turbulent flows}.
\newblock {\em Ann.\ Rev.\ Fluid Mech.}, 44:203--225, 2012.

\bibitem{Khatavkar2006}
V.~V. Khatavkar, P.~D. Anderson, and H.~E.~H. Meijer.
\newblock On scaling of diffuse-interface models.
\newblock {\em Chem. Engng Sci.}, 61:2364--2378, 2006.

\bibitem{Klostermann2013}
J.~Klostermann, K.~Schaake, and R.~Schwarze.
\newblock Numerical simulation of a single rising bubble by vof with surface
  compression.
\newblock {\em Int. J. Numer. Meth. Fluids}, 71:960--982, 2013.

\bibitem{Li1999}
Jie Li and Yuriko Renardy.
\newblock Direct simulation of unsteady axisymmetric core-annular flow with
  high viscosity ratio.
\newblock {\em J.\,Fluid Mech.}, 391:123--149, 1999.

\bibitem{Magaletti2014}
F.~Magaletti, F.~Picano, M.~Chinappi, L.~Marino, and C.~M. Casciola.
\newblock {The sharp-interface limit of the Cahn--Hilliard/Navier--Stokes model
  for binary fluids}.
\newblock {\em J.\,Fluid Mech.}, 714:95--126, 2013.

\bibitem{Mirjalili2018}
Shahab Mirjalili, Christopher~Blake Ivey, and Ali Mani.
\newblock Comparison between the diffuse interface and volume of fluid methods
  for simulating two-phase flows.
\newblock {\em arXiv preprint arXiv:1803.07245}, 2018.

\bibitem{Mitrinovic2000}
D.~Mitrinovic, A.~M. Tikhonov, M.~Li, Z.~Huang, and M.~L. Scholossman.
\newblock Noncapillary wave structure at the water–alkane interface.
\newblock {\em Phys.\ Rev.\ Lett.}, 85:582--585, 2000.

\bibitem{Tryggvason2009}
A.~Prosperetti and G.~Tryggvason.
\newblock {\em Computational methods for multiphase flow}.
\newblock Cambridge University Press, 2009.

\bibitem{rampp2015nscouette}
M~Rampp, J-M Lopez, L~Shi, B~Hof, and M~Avila.
\newblock {NSCOUETTE: A Hybrid MPI-OpenMP Parallel Implementation for
  Pseudospectral Simulations--Scaling experiments on SuperMUC}.
\newblock In {\em 28th International Conference on Supercomputing (ICS 2014)},
  2015.

\bibitem{Roccon2019}
A.~Roccon, F.~Zonta, and A.~Soldati.
\newblock Turbulent drag reduction by compliant lubricating layer.
\newblock {\em J.\,Fluid Mech.}, 863:R1, 2019.

\bibitem{Sahu2016}
K.C. Sahu and R.~Govindarajan.
\newblock Linear stability analysis and direct numerical simulation of two
  layer channel flow.
\newblock {\em J.\,Fluid Mech.}, 798:889--909, 2016.

\bibitem{Scarbolo2015}
L.~Scarbolo, F.~Bianco, and A.~Soldati.
\newblock Coalescence and breakup of large droplets in turbulent channel flow.
\newblock {\em Phys.\ Fluids}, 27:1--6, 2016.

\bibitem{Shi2015}
L.~Shi, M.~Rampp, B.~Hof, and M.~Avila.
\newblock {A hybrid MPI-OpenMP parallel implementation for pseudospectral
  simulations with application to Taylor--€"Couette flow}.
\newblock {\em Computers \& Fluids}, 106:1--11, 2015.

\bibitem{Sussman1994}
M.~Sussman, P.~Smereka, and S.~Osher.
\newblock Free energy of a nonuniform system. i. interface free energy.
\newblock {\em J.\,Comput.\ Phys.}, 114:146--159, 1994.

\bibitem{Tryggvason2001}
G.~Tryggvason, B.~Bunner, A.~Esmaeeli, D.~Juric, N.~AL{-}Rawahi, W.~Tauber,
  J.~Han, S.~Nas, and Y.~J. Jan.
\newblock A front-tracking method for the computations of multiphase flow.
\newblock {\em J.\,Comput.\ Phys.}, 169:708–759, 2001.

\bibitem{willis2013}
A.P. Willis, P.~Cvitanovi\'c, and M.~Avila.
\newblock {Revealing the state space of turbulent pipe flow by symmetry
  reduction}.
\newblock {\em J.\,Fluid Mech.}, 721:514--540, 2013.

\bibitem{openpipeflow}
Ashley~P Willis.
\newblock The {O}penpipeflow {N}avier--{S}tokes solver.
\newblock {\em SoftwareX}, 6:124--127, 2017.

\bibitem{Yue2007}
P.~Yue, C.~Zhou, and J.~J. Feng.
\newblock Spontaneous shrinkage of drops and mass conservation in phase{-}field
  simulations.
\newblock {\em J.\,Comput.\ Phys.}, 223:1--9, 2007.

\bibitem{Yue2010}
P.~Yue, C.~Zhou, and J.~J. Feng.
\newblock {Sharp-interface limit of the Cahn--€"Hilliard model for moving
  contact lines}.
\newblock {\em J.\,Fluid Mech.}, 645:279–294, 2010.

\end{thebibliography}

\end{document}